\documentclass{article}

\usepackage{PRIMEarxiv}

\usepackage[utf8]{inputenc} % allow utf-8 input
\usepackage[T1]{fontenc}    % use 8-bit T1 fonts
\usepackage{hyperref}       % hyperlinks
\usepackage{url}            % simple URL typesetting
\usepackage{booktabs}       % professional-quality tables
\usepackage{amsfonts}       % blackboard math symbols
\usepackage{nicefrac}       % compact symbols for 1/2, etc.
\usepackage{microtype}      % microtypography
\usepackage{lipsum}
\usepackage{fancyhdr}       % header
\usepackage{graphicx}       % graphics
\graphicspath{{media/}}     % organize your images and other figures under media/ folder

\usepackage{amsmath,xcolor,caption,subcaption}
\title{An outlier-resistant $\kappa$-generalized approach \\ for robust physical parameter estimation
}

\author{
  Sérgio Luiz E. F. da Silva \\
  Seismic Inversion and Imaging Group \\
  Fluminense Federal University \\
  Niterói, RJ, Brazil.\\
  \texttt{sergioluizsilva@id.uff.br} \\
   \And
  R. Silva \\
  Department of Theoretical and Experimental Physics\\
  Federal University of Rio Grande do Norte \\
  Natal, RN, Brazil\\
   \And
  Gustavo Z. dos Santos Lima \\
  School of Science and Technology\\
  Federal University of Rio Grande do Norte \\
  Natal, RN, Brazil\\
   \And
  João M. de Araújo \\
  Department of Theoretical and Experimental Physics\\
  Federal University of Rio Grande do Norte \\
  Natal, RN, Brazil\\
   \And
  Gilberto Corso \\
  Department of Biophysics and Pharmacology\\
  Federal University of Rio Grande do Norte \\
  Natal, RN, Brazil\\
}

\begin{document}
\maketitle

\begin{abstract}
In this work we propose a robust methodology to mitigate the undesirable effects caused by outliers to generate reliable physical models. In this way, we formulate the inverse problems theory in the context of Kaniadakis statistical mechanics (or $\kappa$-statistics), in which the classical approach is a particular case. In this regard, the errors are assumed to be distributed according to a finite-variance $\kappa$-generalized Gaussian distribution. Based on the probabilistic maximum-likelihood method we derive a $\kappa$-objective function associated with the finite-variance $\kappa$-Gaussian distribution. To demonstrate our proposal's outlier-resistance, we analyze the robustness properties of the $\kappa$-objective function with help of the so-called influence function. In this regard, we discuss the role of the entropic index ($\kappa$) associated with the Kaniadakis $\kappa$-entropy in the effectiveness in inferring physical parameters by using strongly noisy data. In this way, we consider a classical geophysical data-inverse problem in two realistic circumstances, in which the first one refers to study the sensibility of our proposal to uncertainties in the input parameters, and the second is devoted to the inversion of a seismic data set contaminated by outliers. The results reveal an optimum $\kappa$-value at the limit $\kappa \rightarrow 2/3$, which is related to the best results.
\end{abstract}

% keywords can be removed
\keywords{Kaniadakis $\kappa$-statistics \and Maximum entropy \and Likelihood function \and Inverse problems \and Seismic imaging }

\section{Introduction}
\label{S:1}
    
In many scientific and industrial fields, the determination of (unknown) causes from observed data is essential in finding physical parameters that cannot be directly measured. For example, the computed tomography \cite{biomedicalImagingBertero} is routinely used to detect bone fractures, in which medical imaging is performed by determining the attenuation and absorption (unknown cause) of X-rays that pass through it and are recorded by detectors (observed data). In this regard, the computerized X-ray imaging procedure aims to extract useful information on the physical parameters (model parameters) of bones (or internal organs, blood vessels, among others) from the observed data (X-ray records), in which a physical and/or a mathematical law associates the measured to the model parameters (also known as the forward problem). The solution to this type of problem requires a set of knowledge from various fields such as physics, statistics, mathematics, and scientific computation, which constitute the basic foundations of the inverse problem theory \cite{tarantola2005book}.

The inverse problem theory has been widely employed to describe a variety of complex systems in several fields such as medical imaging \cite{FWI_brain_nature,PhysRevApplied_MedicalUltrasound_Theis_2020}, physics \cite{Berg_2017,InverseProblemsFisica_2020}, geophysics \cite{daSilva_2017_BSSA_mestrado,FWI_EAGEITALY_PLOSONE2020}, engineering \cite{InverseProblemEngenharia2020,Ernani_inverseProb_2021}, among many others \cite{AppliedPhysics_invProb_200,InverseProblemsBiologia_Engl2009,Marsquakes_EPJB_2021}. However, despite successful applications of the inverse problem theory, it is inherently ill-posed since at least one of the following requirements is often violated \cite{hadamard,inverseproblemsbook_illposed_2011}: \textit{(i)} the solution exists, \textit{(ii)} it is unique, and \textit{(iii)} depends continuously on the observed data. 

From a practical point of view, inverse problems are formulated as optimization tasks that aim to match modeled data to the observed data. The objective function to be minimized in the classical approach is based on the least-squares distance between the modeled data and the observed data. In the classical approach, the errors (the difference between modeled and observed data) are assumed to be Gaussian \cite{tarantola2005book}. Although Gaussian errors are very present in nature, non-Gaussian noises are not uncommon in specific systems, so we need to deal with such interferences using a different approach. Indeed, just a handful of outliers is enough to make the classical framework inappropriate \cite{ClaerboutRobustErraticData1973,huberOriginal1973,ParameterEstimation_Constable_1988}. In this way, objective functions based on non-Gaussian criteria are necessary for a robust (and reliable) estimate of physical parameters.

In this way, several robust statistical methods have been developed to solve problems of inferring physical parameters, with particular attention to objective functions that are not severely affected by outliers. In this sense, the literature has demonstrated that non-Gaussian criteria are more suitable for a robust solution of inverse problems. Among them, we can mention the objective functions based on heavy-tailed probability distributions \cite{studentT_tristan_2012,CauchyRegression_2017,Ubaidillah_2017_studentT}, hybrid functions \cite{huberOriginal1973,BubeLangan_1997_hybridObjectiveFunction,SEG2019_Sergio_Shannon}, Laplace distribution \cite{tarantola1984,FWI_qLaplace_2021,PSI_EPJplus_KaniadakisTsallis_2021}, and those based on the generalized Gauss' law of error \cite{q_GaussianPhysicaA_DaSilva_2020,PSI_Jackson_entropy_2021,EPJ_plus_Joao_2021} and deformed maximum likelihood approaches \cite{ferrari_MaxLqLike_2010,q_GaussRobustness_MaxLqLike_japoneses2009,PRE_STM_Kaniadakis_2021}.

Recently, Ref.~\cite{PhysRevE.101.053311_FWI_Kaniadakis} introduced a robust objective function based on a one-parameter generalized Gaussian distribution in the sense of $\kappa$-statistics \cite{KANIADAKIS2001405,KaniadakisRelatividade1,KaniadakisRelatividade2,KaniadakisEntropyReview}. In this regard, the so-called $\kappa$-Gaussian probability distribution was constructed by replacing the ordinary exponential function, $\exp$, with the $\kappa$-deformed exponential function (or $\kappa$-exponential), $\exp_{\kappa}$. Regarding the study reported in Ref.~\cite{PhysRevE.101.053311_FWI_Kaniadakis}, they found that, although $\kappa$-parameter deforms the tail of the $\kappa$-Gaussian distribution, the robustness of the estimates was not significantly affected by the $\kappa$-value. However, despite its successful application to a very complex problem, the robustness properties of data-inverse problems based on the $\kappa$-Gaussian and the $\kappa$-optimum value remains an open question. Moreover, the direct replacement of the ordinary exponential function by the $\kappa$-exponential deforms only (significantly) the distribution's tails and not its inflection points. Nevertheless, the inflection points of a probability distribution are important for including information on how the random variables (in our case, the data-errors) are spread out around the mean.

In this study, we have revisited the inverse problem theory recently introduced in Ref.~\cite{PhysRevE.101.053311_FWI_Kaniadakis} by considering a $\kappa$-Gaussian distribution constrained by a finite variance in the
context of $\kappa$-statistics, which we call, henceforward, finite-variance $\kappa$-Gaussian distribution. Such a constraint allows the $\kappa$-value to control not only the tails of the $\kappa$-distribution, but also the inflection points of the $\kappa$-Gaussian curve. In this way, we introduce a more flexible error law capable of managing both minor errors (through the variability of errors) and spurious measures or outliers (through the heavy tails of the $\kappa$-distribution). In this regard, as will be shown later, our proposal is a powerful tool capable of being applied to a wide range of problems for describing and analyzing various complex systems.

We have organized this paper as follows. In Section \ref{sec:classicalapproach}, we present a brief review of the classical formulation of the inverse problem theory, which is based on standard statistical mechanics.  In particular, we show, from a statistical viewpoint, the sensitivity of the classical approach to spurious measures (outliers). Then, in Section~\ref{sec:metodologia}, we present the theoretical foundations of $\kappa$-statistics to derive the $\kappa$-generalization of the Gaussian distribution constrained by a finite variance. Besides, we formulate the inverse problem theory from the probabilistic maximum-likelihood framework applied to the finite-variance $\kappa$-Gaussian probability distribution. Moreover, we discuss, both analytically and numerically, the determination of the optimal $\kappa$-parameter. In Section~\ref{sec:NumericalExample}, we present applications of our proposal to solve a classical seismic imaging problem, so-called Post-Stack Inversion (PSI). Finally, in Section~\ref{sec:finalremarks}, we discuss the robustness of our proposal and the role of the $\kappa$-parameter in the sense of the estimation of the physical parameters from very noisy data sets.

%%%%%%%%%%%%%%%%%%%%%%%%%%%%%%%%%%%%%%%%%%%%%%%%%%%%%%%%%%%%%%%%%%%%%%%%%%%%%%
%%%%%%%%%%%%%%%%%%%%%%%%%%%%%%%%%%%%%%%%%%%%%%%%%%%%%%%%%%%%%%%%%%%%%%%%%%%%%%
%%%%%%%%%%%%%%%%%%%%%%%%%%%%%%%%%%%%%%%%%%%%%%%%%%%%%%%%%%%%%%%%%%%%%%%%%%%%%%
\section{\label{sec:classicalapproach} Classical inverse problem approach} 
%%%%%%%%%%%%%%%%%%%%%%%%%%%%%%%%%%%%%%%%%%%%%%%%%%%%%%%%%%%%%%%%%%%%%%%%%%%%%%
%%%%%%%%%%%%%%%%%%%%%%%%%%%%%%%%%%%%%%%%%%%%%%%%%%%%%%%%%%%%%%%%%%%%%%%%%%%%%%
%%%%%%%%%%%%%%%%%%%%%%%%%%%%%%%%%%%%%%%%%%%%%%%%%%%%%%%%%%%%%%%%%%%%%%%%%%%%%%

In the classical approach, the inverse problem theory assumes that the errors ($\vec{\varepsilon} = \varepsilon_1, \varepsilon_2, ... , \varepsilon_n$) are independent and identically distributed (\textit{iid}) according to a Gaussian probability distribution \cite{tarantola2005book}:
\begin{equation}
    p(\varepsilon) = Z \exp\Bigg(-\beta \varepsilon^2\Bigg),
\label{eq:standardGaussianDistribution0} 
\end{equation}
where $Z$ is a normalizing constant and $\beta > 0$ is a constant. Usually, the standard Gaussian distribution is considered, which means that the probability function in \eqref{eq:standardGaussianDistribution0} is constrained to both the normalization condition,
\begin{equation}
    \int_{-\infty}^{+\infty} p(\varepsilon) d\varepsilon  = 1,
    \label{eq:NormalizationCondition}
\end{equation}
and the unit second moment (unit variance),
\begin{equation}
    \int_{-\infty}^{+\infty} \varepsilon^2 p(\varepsilon) d\varepsilon  = 1.
    \label{eq:secondMoment}
\end{equation}
We notice that the application of constraints \eqref{eq:NormalizationCondition} and \eqref{eq:secondMoment} in the probability function given in Eq.~\eqref{eq:standardGaussianDistribution0} results in $Z = 1/\sqrt{2\pi}$ and $\beta = 1/2$. In this way, the classical inverse problem assumes that the errors are \textit{iid} according to the well-known standard Gaussian distribution \cite{tarantola2005book}:
\begin{equation}
    p(\varepsilon) = \frac{1}{\sqrt{2\pi}} \exp\Bigg(-\frac{1}{2} \varepsilon^2\Bigg).
\label{eq:standardGaussianDistribution} 
\end{equation}

Since the errors are assumed to be \textit{iid}, the classical inverse problem is formulated through the following optimization problem:
\begin{equation}
    \underset{\mathbf{m}}{max} \hspace{.1cm} \mathcal{L} (\textbf{m}) := \prod_{i=1}^n p\Big(\varepsilon_i(\textbf{m})\Big),
    \label{eq:gaussian_likelihood}
\end{equation}
where $\mathcal{L}$ represents the likelihood function, $n$ is the sample size, and $\textbf{m}$ are the model parameters to be estimated, in which $\varepsilon(\textbf{m}) = d_{mod}(\textbf{m}) - d_{obs}$ with $d_{mod}$ and $d_{obs}$ being the modeled and observed data, respectively. 

The modeled data is computed in the so-called forward problem, in which a physical and/or a mathematical law links the model space to the data space through forward operation: $d_{mod}(\textbf{m}) = G(\textbf{m})$, for $G$ being the forward operator. For instance, in the issue of inferring the gravitational field of the Earth from the gravitational acceleration recorded by gravimeters, the model parameter is the mass distribution in an area of interest, the observed data is the Earth's gravitational acceleration on the surface,  and the forward operator is given by Newton's law of universal gravitation.

It is worth mentioning that the classical objective function is obtained through the principle of maximum likelihood, from the combination of Eqs.~\eqref{eq:standardGaussianDistribution} and \eqref{eq:gaussian_likelihood}. Furthermore, we notice that maximizing the likelihood function [Eq.~\eqref{eq:gaussian_likelihood}] or the log-likelihood are equivalent:
\begin{equation}
    \underset{\mathbf{m}}{max} \hspace{.1cm} \mathcal{L} (\textbf{m}) = \underset{\mathbf{m}}{max} \hspace{.1cm} \ln\Bigg(\mathcal{L} (\textbf{m})\Bigg),
    \label{qGaussian_likelihood}
\end{equation}
since $\mathcal{L}$ is a monotone increasing function of $\ln\big(\mathcal{L}\big)$. Thus:
\begin{equation}
    \ln\Bigg(\mathcal{L} (\textbf{m})\Bigg) = -\frac{n}{2} \ln\Big(2\pi\Big) -\frac{1}{2}\sum_{i=1}^n \varepsilon_i^2(\textbf{m}).
    \label{eq:qGaussian_likelihood_2}
\end{equation}
Since the first term on the right side of the latter equation is constant, the maximization of the likelihood function in \eqref{eq:gaussian_likelihood} is equivalent to the following minimization problem:
\begin{equation}
    \underset{\mathbf{m}}{min} \hspace{.1cm} \phi(\textbf{m}) := \frac{1}{2}\sum_{i=1}^n \varepsilon_i^2(\textbf{m}),
    \label{eq:classicalobjectivefunction}
\end{equation}
in which $\phi$ denotes the classical objective function.

However, the assumption that errors are Gaussian is often inappropriate, due to the existence of outliers in the observations. To see this, it is interesting to analyze the objective function sensitivity to change in the errors' distribution. In this way, we notice that a necessary condition for the objective-function optimality is:
\begin{equation}
     \sum_{i=1}^n \alpha_i \Upsilon = 0,
\end{equation}
where $\alpha_i$ are arbitrary constants and $\Upsilon$ is the influence function  \cite{BookInfluenceFunction_Hampel2005,InfluenceFunction_Hampel1974}:
\begin{equation}
    \Upsilon := \frac{\partial \phi(\textbf{m})}{\partial m_k},
    \label{eq:inflencefunctiondef}
\end{equation}
in which $m_k$ is the $k$-th model parameter. An influence function is a useful tool for analyzing the objective-function robustness regarding erratic observations \cite{ThesisInfluenceFunction_Hampel1968}. In this regard, an objective function is said to be robust (resistant to outliers) if and only if $\Upsilon \rightarrow 0$ under $d^{obs} \rightarrow \infty$ \cite{tarantola2005book}. Otherwise, the function is not robust if $\Upsilon \rightarrow \infty$ under $d^{obs} \rightarrow \infty$.

In this way, the influence function of the classical objective function (hereinafter classical influence function) is given by:
\begin{equation}
    \Upsilon_{\phi} = \sum_{i=1}^n \varepsilon_i(m_k) = \sum_{i=1}^n \Big(d_{{mod}_i}(m_k) - d_{{obs}_i}\Big).
    \label{eq:influclass}
\end{equation}
Note that, facing the existence of outliers ($d^{obs} \rightarrow \infty$), the classical influence function tends to infinity ($\Upsilon_\phi \rightarrow \infty$), which confirms the non-robustness of the classical approach to spurious measures. In this sense, the inverse problems based on Gaussian behaviors are very sensitive to outliers in the data set and, therefore, robust formulations are useful for a reliable estimation of physical parameters of complex systems.

From a statistical viewpoint, the classical approach is very sensitive to outliers because its maximum likelihood estimator (MLE) is related with the observed-data expected value (mean). To illustrate this relationship, suppose that an experiment is performed $n$ times and that each observation $d_{obs_i}$ is recorded. For simplicity, let us assume that the observations are noisy measurements associated each with the same model parameter $m_*$. Furthermore, let us consider a linear inverse problem, $\textbf{d}_{mod} = \textbf{G}\textbf{m}$, with $\textbf{G} = [1, 1, \cdots, 1]^T$ and $\textbf{m} = [m_*]$, where the superscript $T$ denotes the transpose. In addition, since the classical approach assumes that errors are \textit{iid} according to a standard Gaussian distribution [Eq.~\eqref{eq:standardGaussianDistribution}], the respective MLE is given by the maximum likelihood principle \cite{MENKE198479}: 
\begin{equation}
    \frac{\partial \ln\Big(\mathcal{L} (m_*)\Big)}{\partial m_*} = 0.
    \label{eq:principioMLE}
\end{equation}

Now, by considering the definition in the likelihood function in \eqref{eq:gaussian_likelihood} and the probability distribution in \eqref{eq:standardGaussianDistribution}, we have:
\begin{equation}
    \frac{\partial \ln\Big(\mathcal{L} (m_*)\Big)}{\partial m_*} =  \sum_{i=1}^n \Big(m_* - d_{obs_i}\Big) = 0,
\end{equation}
which becomes:
\begin{equation}
    \hspace{.2cm} \hat{m}_* = \frac{1}{n}\sum_{i=1}^n d_{obs_i},
\end{equation}
where $\hat{m}_*$ is the estimated value of $m_*$. Indeed, as can be seen in the latter equation, the MLE associated with the classical inverse problem is the observed-data mean [see Ref.~\cite{MENKE198479} for more details]. In this regard, if there is a handful of spurious measures, for instance, the classical approach fails to estimate the model parameters because the expected value is very sensitive to outliers.

In addition, we notice that the statistical analysis performed using the principle of maximum likelihood [Eq.~\eqref{eq:principioMLE}] is equivalent to evaluate the classical influence function [Eq.~\eqref{eq:influclass}] at its stationary point, $\Upsilon_{\phi}=0$:
\begin{equation}
    \Upsilon_{\phi} = \sum_{i=1}^n \varepsilon_i(m_k) = \sum_{i=1}^n \Big(m_* - d_{obs_i}\Big) = 0.
\end{equation}

%%%%%%%%%%%%%%%%%%%%%%%%%%%%%%%%%%%%%%%%%%%%%%%%%%%%%%%%%%%%%%%%%%%%%%%%%%%%%%
%%%%%%%%%%%%%%%%%%%%%%%%%%%%%%%%%%%%%%%%%%%%%%%%%%%%%%%%%%%%%%%%%%%%%%%%%%%%%%
%%%%%%%%%%%%%%%%%%%%%%%%%%%%%%%%%%%%%%%%%%%%%%%%%%%%%%%%%%%%%%%%%%%%%%%%%%%%%%
\section{\label{sec:metodologia} Robust inverse problem based on $\kappa$-Gaussian distributions} 
%%%%%%%%%%%%%%%%%%%%%%%%%%%%%%%%%%%%%%%%%%%%%%%%%%%%%%%%%%%%%%%%%%%%%%%%%%%%%%
%%%%%%%%%%%%%%%%%%%%%%%%%%%%%%%%%%%%%%%%%%%%%%%%%%%%%%%%%%%%%%%%%%%%%%%%%%%%%%
%%%%%%%%%%%%%%%%%%%%%%%%%%%%%%%%%%%%%%%%%%%%%%%%%%%%%%%%%%%%%%%%%%%%%%%%%%%%%%

In recent years, the Kaniadakis statistics (or $\kappa$-statistics) \cite{KANIADAKIS2001405} have been widely applied from the $\kappa$-entropy definition, which is a generalization of the classic Boltzmann-Gibbs-Shannon (BGS) entropy. Such framework is based on the $\kappa$-entropy given by:
\begin{equation}
    \mathcal{S}_\kappa [f] := -\sum_{i=1}^n f(x_i) \ln_\kappa\Big(f(x_i)\Big),
    \label{eq:KappaEntropy}
\end{equation}
where $x$ is a random variable, $f$ represents a probability function, and 
\begin{equation}
    \ln_\kappa (y) = \frac{y^{\kappa}-y^{-\kappa}}{2\kappa}
    \label{eq:KappaLogaritmo}
\end{equation}
is known as $\kappa$-logarithm function \cite{KaniadakisRelatividade1,KaniadakisRelatividade2}, for $|\kappa| < 1$ the entropic index. It is worth noting that at limit $\kappa \rightarrow 0$, the $\kappa$-logarithmic [Eq.~\eqref{eq:KappaLogaritmo}] is reduced to the usual logarithmic function, $\ln(y)$, and therefore the BGS entropy is recovered in the same limit:
\begin{equation}
    \lim_{\kappa\to 0} \mathcal{S}_\kappa [f] = -\sum_{i=1}^n f(x_i) \ln\Big(f(x_i)\Big) = \mathcal{S}_{BGS} [f].
    \label{eq:BGEntropy}
\end{equation}
 
In the same way that the natural logarithm, $\ln(y)$, has the exponential function, $\exp(y)$,  as the inverse function, the $\kappa$-logarithmic is the inverse of the $\kappa$-exponential function, $\exp_\kappa(y)$: 
\begin{equation}
    \ln_\kappa\Big(\exp_\kappa (y) \Big) = \exp_\kappa\Big(\ln_\kappa (y) \Big) = y,
\end{equation}
with
\begin{equation}
    \exp_\kappa (y) = exp\Bigg(\frac{1}{\kappa} arcsinh(\kappa x)\Bigg) = 
    \Big(\sqrt{1+\kappa^2 y^2}+\kappa y \Big)^\frac{1}{\kappa}.
    \label{eq:kappaExponential}
\end{equation}
It is noteworthy that these $\kappa$-functions [Eqs.~\eqref{eq:KappaLogaritmo} and \eqref{eq:kappaExponential}] have been applied in a wide variety of studies and direct applications of this formalism have been carried out in different contexts, such as in quantum physics \cite{quantumHtheoremRaimundoDoryHelio_2011,quantumKaniadakis_Ourabah_2015,quantumKaniadakis_SOARES_2019}, astrophysics \cite{renanKaniadakis,KaniadakisAppAstrofisica,KaniadakisAppAstrofisica2_amelia2015}, complex networks \cite{ComplexNetworksKaniadakis1,ComplexNetworksKaniadakis2,DASILVA2021125539_NewtonCoolingLaw}, epidemiology \cite{KaniadakisEpidemology2020}, power-law distributions \cite{Kaniadakis_MaxEnt_PowerLawDistribution_2009,Kaniadakis_NewPowerLawDistribution_2021}, as well as in geophysics \cite{PhysRevE.89.052142,PhysRevE.89.052142__2__entropy,kappaGRlaw_daSilva_chaosSolitionsFractals_2021}.

In this work, we use the $\kappa$-exponential function, and therefore it is worth mentioning some of its properties. Since we are interested in analyzing the effects of the outliers on parameters inference, one of the most important characteristic of $exp_\kappa(y)$ function is its asymptotic behavior \cite{KaniadakisEntropyReview}:
\begin{equation}
    \lim_{y \to \pm \infty} \exp_\kappa(y) \sim |2\kappa y|^{\pm 1/\kappa}.
    \label{eq:comportamentoAssintoticoKappaExp}
\end{equation}

Based on the $\kappa$-exponential function [Eq.~\eqref{eq:kappaExponential}], Ref.~\cite{PhysRevE.101.053311_FWI_Kaniadakis} formulated a complex inverse problem assuming that the errors are \textit{iid} according to the following $\kappa$-generalization of the Gaussian distribution (hereinafter traditional $\kappa$-Gaussian distribution):
\begin{equation}
    p_\kappa(\varepsilon) = C_\kappa \exp_\kappa\Bigg(-\frac{1}{2}\varepsilon^2\Bigg),
    \label{eq:kappaGaussian_trad}
\end{equation}
where 
\begin{equation}
    C_\kappa = \big(1+2|\kappa|\big)\sqrt{\frac{|\kappa|}{\pi}}\hspace{.1cm}\frac{\Gamma\Big(\frac{1}{|2\kappa|}+\frac{1}{4}\Big)}{\Gamma\Big(\frac{1}{|2\kappa|}-\frac{1}{4}\Big)}
\end{equation}
is the normalization factor. Although the inverse problems based on the traditional $\kappa$-Gaussian distribution [Eq.~\eqref{eq:kappaGaussian_trad}] have been quite successful in damping spurious values (outliers), the statistical constraints in \eqref{eq:NormalizationCondition} and \eqref{eq:secondMoment} were not taken into account. In other words, the dispersion of the error was not considered. However, dispersion is an important factor since it brings information about how a given probability distribution is squeezed or stretched \cite{devore2011}.

\begin{figure}[!htb]
\flushleft{\hspace{.6cm} (a) \hspace{7.5cm} (b)}
    \includegraphics[width=\textwidth]{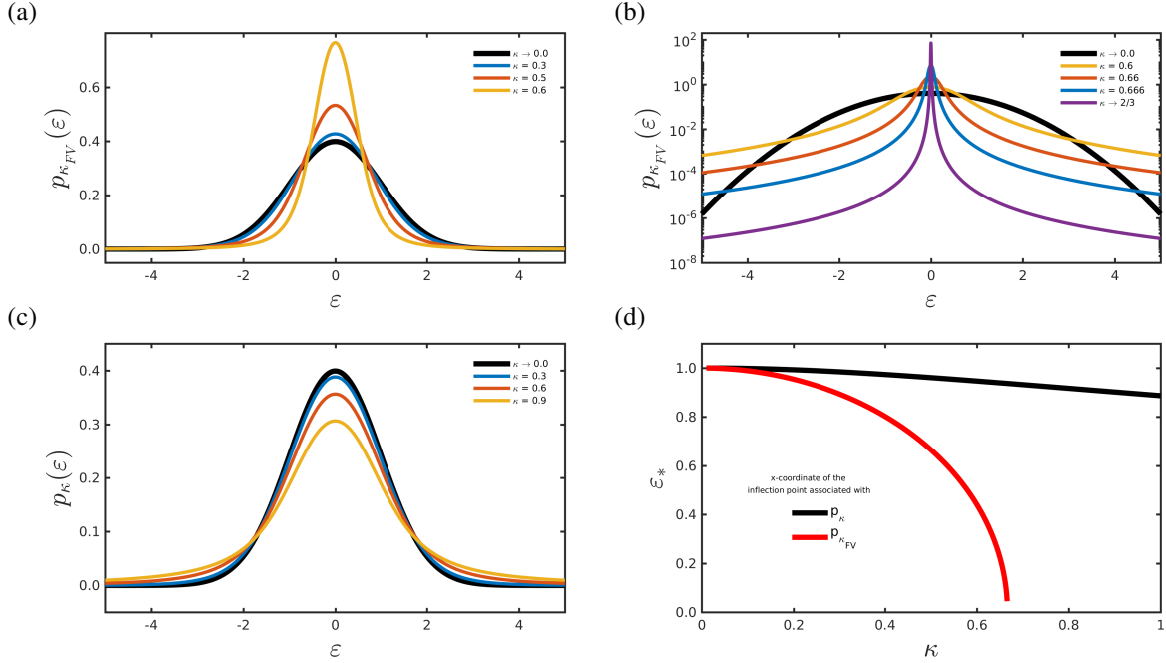}
\flushleft{\vspace{-5.5cm}\hspace{.6cm} (c) \hspace{7.5cm} (d)}
\vspace{5cm}
    \caption{(a)-(b) Finite-variance $\kappa$-Gaussian distribution curves for typical $\kappa$-values [Eq.~\eqref{eq:kappaGaussian_new_general0}], in which the panel (b) shows the finite-variance  $\kappa$-probability semi-log plots for the $\kappa \rightarrow 2/3$ limit case. (c) Traditional $\kappa$-Gaussian probability plots [Eq.~\eqref{eq:standardGaussianDistribution}]. The black curve in panels (a)-(c) represents the standard Gaussian distribution ($\kappa \rightarrow 0$).  Panel (d) indicates the $x$-coordinate of the inflection point [Eq.~\eqref{eq:inflexionpoint}] associated with the entropic index $\kappa$, in which the black curve refers to the traditional $\kappa$-probability function %proposed by Ref.~\cite{PhysRevE.101.053311_FWI_Kaniadakis}
     [Eq.~\eqref{eq:standardGaussianDistribution}], and the red curve relates to the finite-variance $\kappa$-probability distribution [Eq.~\eqref{eq:kappaGaussian_new_general0}].}
    \label{fig:KappaGaussianProbabilityPlots}
\end{figure}

In summary, the most significant differences among the traditional $\kappa$-Gaussian distributions according to the entropic index $\kappa$ are in the tails of the distribution [see Fig.~\ref{fig:KappaGaussianProbabilityPlots}(c)]. In this way, independent of the $\kappa$-value, the function has heavy tails, and therefore the referred approach proved to be robust regardless of the $\kappa$-value. However, this approach does not take into account the variability of the errors, which means that if there is numerous spurious measures (outliers), the approach based on the traditional $\kappa$-Gaussian distribution is predestined to fail. Furthermore, the optimal $\kappa$-value remains veiled.

In this way, to provide more flexibility both in the tails and in the inflection points of the $\kappa$-Gaussian probability function, we derive in this work an alternative $\kappa$-Gaussian distribution (hereinafter finite-variance $\kappa$-Gaussian distribution), $p_{\kappa_{{}_{FV}}}$, that takes into account the statistical moments [Eqs.~\eqref{eq:NormalizationCondition} and \eqref{eq:secondMoment}]. For this, let us assume that the errors ($\vec{\varepsilon} = \varepsilon_1, \varepsilon_2, ... , \varepsilon_n$) are \textit{iid} according to the following $\kappa$-Gaussian probability distribution:
\begin{equation}
    p_{\kappa_{{}_{FV}}}(\varepsilon) = Z_\kappa \exp_\kappa\Bigg(-\beta_\kappa\varepsilon^2\Bigg),
    \label{eq:kappaGaussian_new_general0}
\end{equation}
where $Z_\kappa > 0$ is a normalizing constant and $\beta_\kappa > 0$ depends on the entropic index $\kappa$.

From the normalization condition [Eq.~\eqref{eq:NormalizationCondition}], the normalizing constant can be computed by:
\begin{equation}
    \frac{1}{Z_\kappa} = \int_{-\infty}^{+\infty} \exp_\kappa\Bigg(-\beta_\kappa\varepsilon^2\Bigg) d\varepsilon, 
    \label{eq:zkappaapp}
\end{equation}
which, after some algebra, results in:
\begin{equation}
    Z_\kappa = \Bigg(1+\frac{|\kappa|}{2}\Bigg)\frac{\Gamma\Big(1/|2\kappa|+1/4\Big)}{\Gamma\Big(1/|2\kappa|-1/4\Big)} \sqrt{\frac{2|\kappa|\beta_\kappa}{\pi}},
    \label{zkappa}
\end{equation}
where $\Gamma(y) = \int_0^\infty t^{y-1} exp(-t) dt$ is the Euler gamma function. The $\beta_\kappa$-parameter is determined by applying the constraint associated with the unit second moment [Eq.~\eqref{eq:secondMoment}]:
\begin{equation}
    \int_{-\infty}^{+\infty} \varepsilon^2 Z_\kappa \exp_\kappa\Bigg(-\beta_\kappa\varepsilon^2\Bigg) d\varepsilon= 1.
    \label{eq:integralxkappa2}
\end{equation}
By solving the integral in Eq.~\eqref{eq:integralxkappa2} and considering the normalization constant in Eq.~\eqref{zkappa}, $\beta_\kappa$-parameter is given by:
\begin{equation}
    \beta_\kappa = \frac{(1+|\kappa|/2)}{|2\kappa|(2+3|\kappa|)} \frac{\Gamma(1/|2\kappa|-3/4)\Gamma(1/|2\kappa|+1/4)}{\Gamma(1/|2\kappa|+3/4)\Gamma(1/|2\kappa|-1/4)},
    \label{eq:betadefinicao}
\end{equation}
which is valid for $|\kappa| < 2/3$ so that conditions $Z_\kappa > 0$ and $\beta_\kappa > 0$ are satisfied. In addition, it is worth emphasizing that the integrals solution in Eqs.~\eqref{eq:zkappaapp} and \eqref{eq:integralxkappa2} were carried out with the help of Eq.~(A20) of Ref.~\cite{KaniadakisRelatividade2}.

Thus, the finite-variance $\kappa$-Gaussian distribution is given by Eq.~\eqref{eq:kappaGaussian_new_general0} with $Z_\kappa$ and $\beta_\kappa$ given by Eqs.~\eqref{zkappa} and \eqref{eq:betadefinicao}, respectively. Furthermore, it is worth noting that in the $\kappa \rightarrow 0$ classic limit the finite-variance $\kappa$-Gaussian distribution in \eqref{eq:kappaGaussian_new_general0} reduces to the standard Gaussian distribution, since 
\begin{equation}
    \lim_{\kappa \to 0} \beta_\kappa = \frac{1}{2} = \beta \qquad\mbox{and}\qquad \lim_{\kappa \to 0} Z_\kappa = \frac{1}{\sqrt{2\pi}} = Z.
    \label{eq:betavalor}
\end{equation}
Furthermore, we notice that the traditional $\kappa$-Gaussian distribution is a particular case of the finite-variance $\kappa$-Gaussian distribution when $\beta_\kappa = 1/2$.

Figure~\ref{fig:KappaGaussianProbabilityPlots} shows the probability plots of the traditional and finite-variance  $\kappa$-Gaussian distributions for typical $\kappa$-values, in which the standard Gaussian distribution ($\kappa \rightarrow 0$) is depicted by the black curve in the panels (a)-(c). In this regard, the panels (a) and (b) depict the finite-variance  $\kappa$-Gaussian distributions curves [Eq.~\eqref{eq:kappaGaussian_new_general0}], in which the panel (b) shows a $\kappa$-probability semi-log plots for the $\kappa \rightarrow 2/3$ limit case. We notice that in the particular case $\kappa \rightarrow \frac{2}{3}^-$. 
The traditional $\kappa$-Gaussian distribution
[Eq.~\eqref{eq:standardGaussianDistribution}] is depicted in Fig.~\ref{fig:KappaGaussianProbabilityPlots}(c). 

From a visual inspection of Figs.~\ref{fig:KappaGaussianProbabilityPlots}(a) and \ref{fig:KappaGaussianProbabilityPlots}(c) it is remarkable that imposing that the second moment exists, being finite and unitary [Eq.~\eqref{eq:secondMoment}], impacts both the tails and the inflection points of the probability functions. In the following we calculate the inflection point of the finite-variance $\kappa$-Gaussian distribution [Eq.~\eqref{eq:kappaGaussian_new_general0}]. Consider the $\varepsilon$-value, $\varepsilon_*$, in which the second derivative of function in \eqref{eq:kappaGaussian_new_general0} vanishes, we have the following expression, after some algebra:
\begin{equation}
    |\varepsilon_{*}| = \Bigg[\beta_\kappa^2\Bigg(2+\kappa^2+2\sqrt{2\kappa^2+1}\Bigg)\Bigg]^{-\frac{1}{4}}.
    \label{eq:inflexionpoint}
\end{equation}
It is worth emphasizing that in the $\kappa \rightarrow 0$ classic limit case, the $x$-coordinate of the inflection point is equal to $1$, due to the assumption in the Eq.~\eqref{eq:secondMoment}. Furthermore, note that regardless of the $\kappa$-value for the case of the traditional $\kappa$-Gaussian distribution [$\beta_\kappa = 1/2$], the variations in $x$-coordinate value of the inflection point are minimal [see black curve in Fig.~\ref{fig:KappaGaussianProbabilityPlots}(d)] compared to our proposal [see red curve in Fig.~\ref{fig:KappaGaussianProbabilityPlots}(d)].

It is also worth mentioning that, in contrast to the classical approach, error laws based on finite-variance $\kappa$-Gaussian distributions benefit from heavy tails of this distribution, depending on the $\kappa$-entropic index, which decays asymptotically as a power-law: $p_{\kappa_{{}_{FV}}} \big(|\varepsilon| \big) \sim |2\kappa \beta_\kappa \varepsilon^2|^{\pm 1/\kappa}$ [see Eq.~\eqref{eq:comportamentoAssintoticoKappaExp}].

By assuming that the errors ($\vec{\varepsilon} = \varepsilon_1, \varepsilon_2, ... , \varepsilon_n$) are \textit{iid} according to the finite-variance $\kappa$-Gaussian probability distribution [Eq.~\eqref{eq:kappaGaussian_new_general0}], the inverse problem based on our proposal is formulated through the following optimization task:
\begin{equation}
    \underset{\mathbf{m}}{max} \hspace{.1cm} \mathcal{L}_{\kappa_{{}_{FV}}} (\textbf{m}) := \prod_{i=1}^n p_{\kappa_{{}_{FV}}}\Big(\varepsilon_i(\textbf{m})\Big),
    \label{eq:kappa_gaussian_likelihood}
\end{equation}
where $\mathcal{L}_{\kappa_{{}_{FV}}}$ represents the $\kappa$-likelihood function.

In the sequence, the $\kappa_{{}_{FV}}$-objective function is derived from the log-$\kappa$-likelihood:
\begin{equation}
    \ln\Bigg(\mathcal{L}_{\kappa_{{}_{FV}}} (\textbf{m})\Bigg) = n \ln\Big(Z_\kappa\Big) +\sum_{i=1}^n \ln\Bigg[\exp_\kappa\Bigg(-\beta_\kappa \varepsilon_i^2(\textbf{m})\Bigg)\Bigg].
    \label{eq:log_kappa_gaussian_likelihood_2}
\end{equation}
Since $n \ln(Z_\kappa)$ is constant, we notice that to maximize Eq.~\eqref{eq:log_kappa_gaussian_likelihood_2} is equivalent to minimize the negative of the term on the right side of Eq.~\eqref{eq:log_kappa_gaussian_likelihood_2}:
\begin{equation}
    \underset{\mathbf{m}}{min} \hspace{.1cm} \phi_{\kappa_{{}_{FV}}}(\textbf{m}) := -\sum_{i=1}^n \ln\Bigg[\exp_\kappa\Bigg(-\beta_\kappa \varepsilon_i^2(\textbf{m})\Bigg)\Bigg],
    \label{eq:kappaObjectiveFunction}
\end{equation}
in which $\phi_{\kappa_{{}_{FV}}}$ denotes the $\kappa_{{}_{FV}}$-objective function.

In the same way as the finite-variance $\kappa$-Gaussian distribution proposed in the present work [Eq.~\eqref{eq:kappaGaussian_new_general0}] is a generalization of the standard Gaussian distribution [Eq.~\eqref{eq:standardGaussianDistribution}], the $\kappa_{{}_{FV}}$-objective function [Eq.~\eqref{eq:kappaObjectiveFunction}], is also a generalization of the classical objective function [Eq.~\eqref{eq:classicalobjectivefunction}]:
\begin{equation}
\lim_{\kappa \to 0} \phi_{\kappa_{{}_{FV}}}(\textbf{m}) = - \sum_{i=1}^n \ln\Bigg[\exp\Bigg(-\beta \varepsilon_i^2(\textbf{m})\Bigg)\Bigg] = \frac{1}{2} \sum_{i=1}^n \varepsilon_i^2(\textbf{m}) = \phi (\textbf{m}),
\end{equation}
in which $\beta = 1/2$ [see Eq.~\eqref{eq:betavalor}].

\begin{figure}[!htb]
\flushleft{\hspace{.6cm}(a) \hspace{7.5cm} (b)}
    \includegraphics[width=\textwidth]{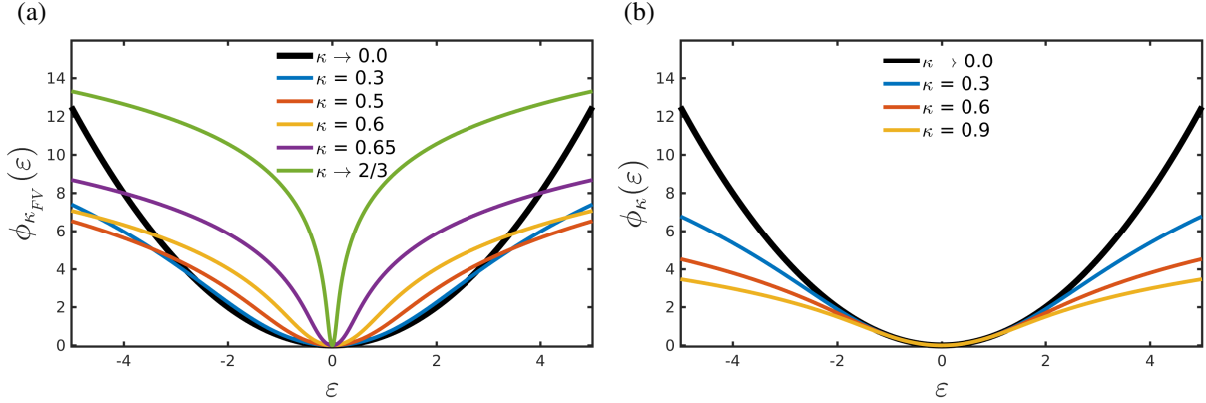}
    \caption{(a) $\kappa_{{}_{FV}}$-objective and (b) traditional $\kappa$-objective functions for typical $\kappa$-values. The black curve represents the classical objective function [Eq.~\eqref{eq:classicalobjectivefunction}].}
    \label{fig:KappaGaussianObjectiveFunction}
\end{figure}

Figure \ref{fig:KappaGaussianObjectiveFunction} shows the behavior of the $\kappa_{{}_{FV}}$-objective function relative to the error, $\varepsilon$, in which the panel (b) depicts the $\beta_\kappa = 1/2$ case, which is related to the traditional $\kappa$-Gaussian distribution. In this figure, the black curve represents the classical case ($\kappa \rightarrow 0$) [Eq.~\eqref{eq:classicalobjectivefunction}]. We notice that the $\kappa_{{}_{FV}}$-objective function [Fig.~\ref{fig:KappaGaussianObjectiveFunction}(a)], as well as the traditional $\kappa$-objective function  [Fig.~\ref{fig:KappaGaussianObjectiveFunction}(b)], down-weight the large errors. Nevertheless, in contrast to the other approaches, the $\kappa_{{}_{FV}}$-objective function up-weight the small error-amplitudes, especially in the $\kappa \rightarrow 2/3$ limit case, as depicted by the green curve in Fig.~\ref{fig:KappaGaussianObjectiveFunction}(a).

To analyze the $\kappa_{{}_{FV}}$-objective function robustness to erratic observations, especially outliers, the $\kappa_{{}_{FV}}$-influence function is defined by [see Eq.~\eqref{eq:inflencefunctiondef}]:
\begin{equation}
    \Upsilon_{\phi_{\kappa_{{}_{FV}}}} := \frac{\partial \phi_{\kappa_{{}_{FV}}}(\textbf{m})}{\partial m_k}.
\end{equation}
In this way, the $\kappa_{{}_{FV}}$-influence function is given by:
\begin{equation}
    \Upsilon_{\phi_{\kappa_{{}_{FV}}}} = \sum_{i=1}^n \frac{2\beta_\kappa \varepsilon_i(m_k)}{\sqrt{1+\kappa^2\beta_\kappa^2\varepsilon_i^4(m_k)}},
    \label{eq:influkappa}
\end{equation}
in which $\varepsilon_i(m_k) = d_{{mod}_i}(m_k) - d_{{obs}_i}$. As expected, in the limit case $\kappa \rightarrow 0$, the $\kappa_{{}_{FV}}$-influence function [Eq.~\eqref{eq:influkappa}] is reduced to the classical influence function [Eq.~\eqref{eq:influclass}]. However, in contrast to the classical approach, the $\kappa_{{}_{FV}}$-objective function is robust for any value of $0 < |\kappa| < 2/3$, since $\Upsilon_{\phi_{\kappa_{{}_{FV}}}}$ tends to zero under $d_{out}^{obs} \rightarrow \infty$ limit-case. It is worth noting that the influence function associated with the traditional $\kappa$-approach, $\Upsilon_{\phi_{\kappa}}$, is determined from the influence function in \eqref{eq:influkappa} with $\beta_\kappa = 1/2$:
\begin{equation}
    \Upsilon_{\phi_{\kappa}} = \sum_{i=1}^n \frac{\varepsilon_i(m_k)}{\sqrt{1+\frac{1}{4}\kappa^2\varepsilon_i^4(m_k)}},
    \label{eq:influkappaPRE}
\end{equation}
which is also resistant to outliers since $\Upsilon_{\phi_{\kappa}} \rightarrow 0$ under $d_{out}^{obs} \rightarrow \infty$. 

\begin{figure}[!htb]
\flushleft{\hspace{.2cm}(a) \hspace{3.5cm} (b) \hspace{3.5cm} (c) \hspace{3.5cm} (d)}
    \includegraphics[width=\textwidth]{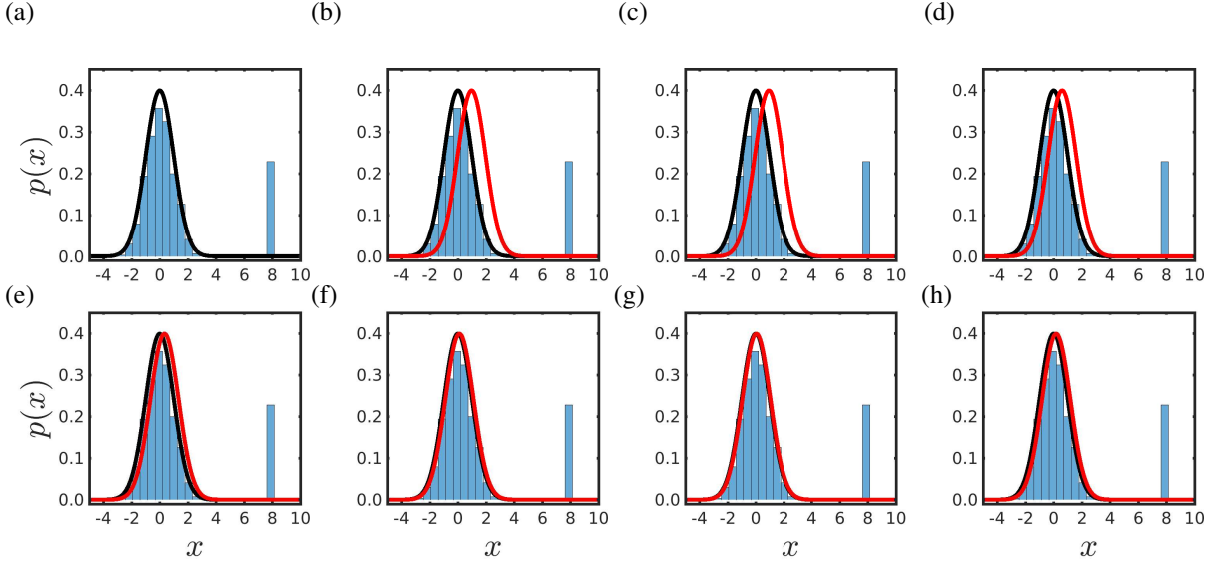}
    \flushleft{\vspace{-4.3cm} \hspace{.2cm}(e) \hspace{3.5cm} (f) \hspace{3.5cm} (g) \hspace{3.5cm} (h) \vspace{3.5cm}}
    \caption{Numerical experiments result to analyze the robustness of inverse problems based on the $\kappa_{{}_{FV}}$-objective function [Eq.~\eqref{eq:kappaObjectiveFunction}], which consists of determining the mean of a standard normal distribution. The data histogram represents the data set contaminated by the outliers at x = 8. The black curve depicts the standard normal distribution [true distribution]. The red curves represent the estimates performed through (b) the classical framework [$\kappa \rightarrow 0$] and based on our proposal with (c) $\kappa = 0.01$, (d) $0.05$, (e) $0.10$, (f) $0.30$, (g) $0.66$ and (h) $k=0.666667$.}
    \label{fig:GaussExperimentRobustness}
\end{figure}

To illustrate the robustness of the $\kappa_{{}_{FV}}$-objective function [Eq.~\eqref{eq:kappaObjectiveFunction}] in relation to outliers, we consider a simple parameter estimation task that consists of inferring the central tendency measures of a standard normal distribution. To perform this experiment, we generate $1,000$ random numbers according to a normal distribution with zero-mean ($\mu = 0$) and unit variance ($\sigma^2 = 1$). Then we contaminate $10\%$ of the sample by introducing outliers at $x = 8$. The standard normal distribution (true distribution) is depicted in Fig.~\ref{fig:GaussExperimentRobustness} by the solid black curve. In this figure, the histogram represents the data set contaminated by the outliers. As predicted by the influence function in Eq.~\eqref{eq:influclass}, the classical approach failed to estimate the central tendency measure of the standard normal distribution [see the discrepancy between the red and black curves in Fig.~\ref{fig:GaussExperimentRobustness}(b)], which confirms the sensitivity of this approach to outliers. The red curves in panels (c)-(h) in Fig.~\ref{fig:GaussExperimentRobustness} represent the estimates performed with our proposal. For cases in which the function behaves similarly to the classical approach, the estimate of the $\mu$-parameter was biased as may be seen by the discrepancy between the red and black curves in the panels (c)-(e), for the cases $\kappa = 0.01$, $0.05$ and $0.10$, respectively. Figure ~\ref{fig:GaussExperimentRobustness} also shows that as the entropic index $\kappa$ increases, the estimated curve [red curves] approaches the true curve [black curves]. Indeed, the $\mu$-parameter estimation, $\hat{\mu}$, for instance $\kappa=0.3$, provides a good estimation as illustrated by the overlapping between the red and black curves [see Fig.~\ref{fig:GaussExperimentRobustness}(f)].

\begin{figure}[!htb]
\flushleft{\hspace{.2cm}(a) \hspace{9.5cm} (b)}
    \includegraphics[width=\columnwidth]{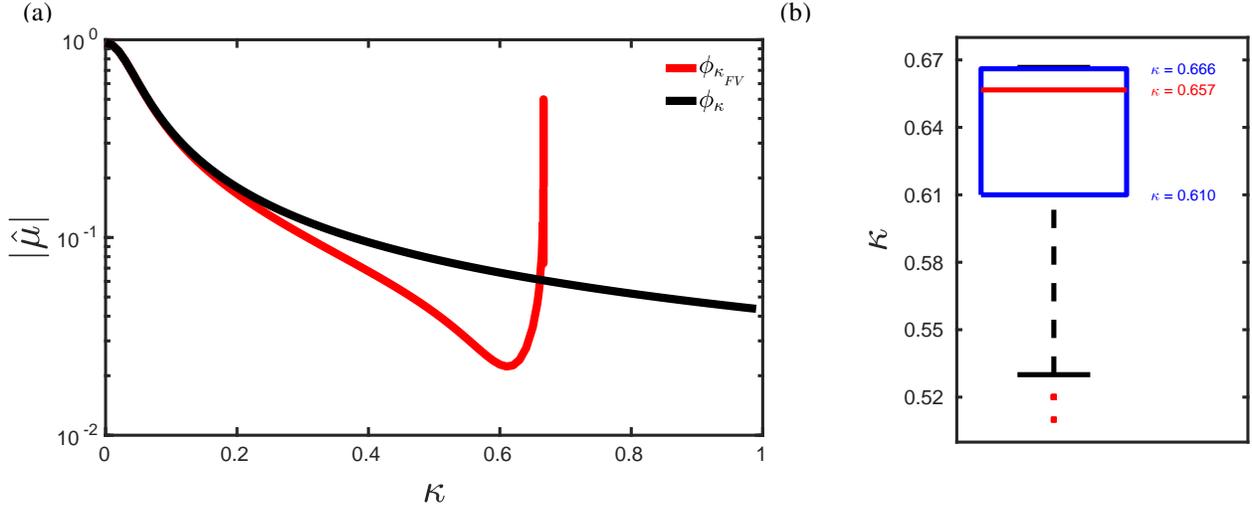}
    \caption{(a) Analysis of robustness in data-inverse for estimating of Normal mean by our proposal [red curve] and the traditional $\kappa$-objective function [black curve]. The best result [$|\hat{\mu}| \rightarrow 0$] is related to our proposal with $\kappa = 0.619$. (b) The boxplot resuming the $\kappa$-values that provided the best estimate of $\mu$ by considering 1,000 experiments.}
    \label{fig:GaussExperimentRobustness_full}
\end{figure}

By considering this same experiment, we perform a simulation for different $\kappa$-values for both our proposal and the traditional $\kappa$-objective function. In this regard, the best result ($|\hat{\mu}| \rightarrow 0$) is associated with the $\kappa_{{}_{FV}}$-objective function [Eq.~\eqref{eq:kappaObjectiveFunction}] with $\kappa= 0.619$, which is represented by the minimal of the red curve in Fig.~\ref{fig:GaussExperimentRobustness_full}(a). Moreover, in order to validate the present example, we performed $1,000$ experiments and determined which value of $\kappa$ provides the best estimate of $\mu$ ($|\hat{\mu}| \rightarrow 0$). In all cases, our proposal overcomes the other approaches presented in this study. The optimal $\kappa$-values found for all numerical experiments are summarized in the boxplot of Fig.~\ref{fig:GaussExperimentRobustness_full}(b), where our proposal outperformed the other approaches presented in this study with a median equals $0.657$, and the first and second quartiles equal $0.610$ and $0.666$, respectively.

The latter experiment motivated the study of the behavior of the $\kappa_{{}_{FV}}$-objective function in the limit $\kappa \rightarrow 2/3$. Indeed, the $\kappa_{{}_{FV}}$-influence function [Eq.~\eqref{eq:influkappa}] in the $\kappa \rightarrow 2/3$ limit-case has an interesting behavior:
\begin{equation}
    \underset{\kappa \rightarrow \frac{2}{3}^-}{ lim } \hspace{.1cm} \Upsilon_{\phi_{\kappa_{{}_{FV}}}} = \sum_{i=1}^n \frac{3}{\varepsilon_i(m_k)} = 3 \sum_{i=1}^n \frac{ sign\Big(\varepsilon_i(m_k)\Big) }{\Big|\varepsilon_i(m_k)\Big|} = \Upsilon_{\phi_{\kappa_{2/3}}},
    \label{eq:upsilon23}
\end{equation}
where the $sign(\varepsilon)$ equals $-1$ for $\varepsilon < 0$, $0$ if $\varepsilon = 0$ and $1$ for $\varepsilon > 0$.

At first glance, the influence function in \eqref{eq:upsilon23} does not provide useful information. For this reason, before discussing the importance of the result reported in Eq.~\eqref{eq:upsilon23}, we analyze a similar influence function, at the stationary point, such as:
\begin{equation}
    \Upsilon_{sign} = \sum_{i=1}^n sign\Big(d_{{mod}_i}(m_k) - d_{{obs}_i}\Big) = 0,
    \label{eq:influencefunctionL1}
\end{equation}
with $\varepsilon_i(m_k) = d_{{mod}_i}(m_k) - d_{{obs}_i}$. In order for the summation in Eq.~\eqref{eq:influencefunctionL1} to be satisfied, $m_k$ should be determined so that half the $d_{{obs}_i}$'s are greater than $\bar{d}_{obs} = \frac{1}{n}\sum_{i=1}^n d_{{obs}_i}$ and the half remaining are smaller than $\bar{d}_{obs}$. In other words, the central element of the observed data set must be chosen, which is exactly the observed-data median [see, for instance, chapter 3 of Ref.~\cite{MENKE198479}]. 

It is worth remembering that, in contrast with the expected value, the median is not skewed by a handful of spurious measures, and therefore it is a robust central tendency measure. The influence function in \eqref{eq:influencefunctionL1} is linked to the objective function based on the sum of the absolute values of the errors ($l_1$-norm of the errors) \cite{tarantola2005book}:
\begin{equation}
    \phi_{sign} = \sum_{i=1}^n \Big|d_{{mod}_i}(m_k) - d_{{obs}_i}\Big|.
\end{equation}
In summary, the MLE associated to the objective function defined in the latter equation is the observed-data median. Although robust objective functions such as defined in the latter equation are desirable, they are not differentiable at $\varepsilon = 0$, which can make difficult the treatment of inverse problems \cite{tarantola2005book}. 

By comparing $\Upsilon_{\phi_{2/3}}$ [Eq.~\eqref{eq:upsilon23}] with $\Upsilon_{sign}$ [Eq.~\eqref{eq:influencefunctionL1}], we notice that the MLE associated to the $\kappa_{{}_{FV}}$-influence function in the $\kappa \rightarrow 2/3$ limit-case is a weighted median \cite{ClaerboutRobustErraticData1973}, in which the weights are proportional to the inverse of the absolute value of the residual data. However, at the limit $\kappa \rightarrow 2/3$, the $\kappa_{{}_{FV}}$-objective function is also not differentiable. In this way, to mitigate the singularity of the influence function $\Upsilon_{\phi_{2/3}}$, we propose a $\kappa$-entropic index close to $(2/3)^-$, instead of exactly $2/3$.

\begin{figure}[!htb]
\flushleft{\hspace{.2cm}(a) \hspace{7.8cm} (b)}
    \includegraphics[width=\textwidth]{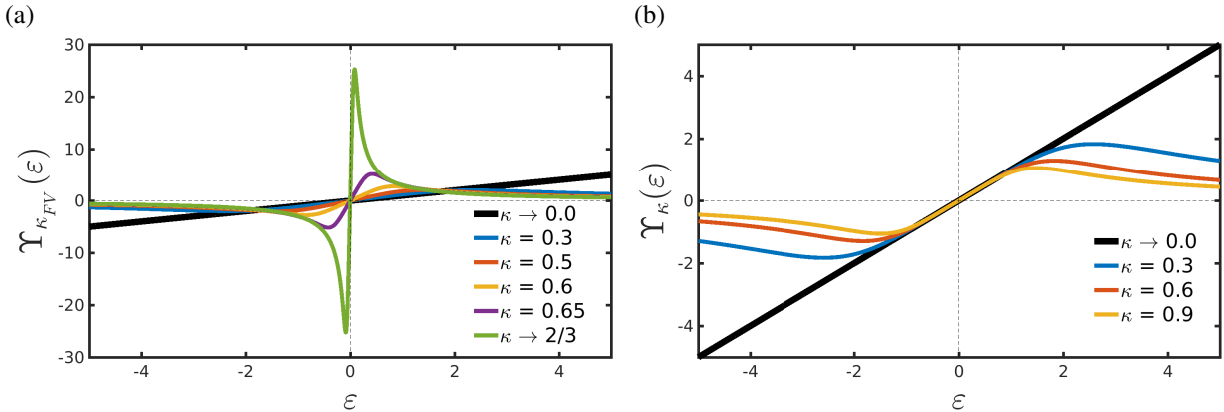}
    \caption{(a) $\kappa_{{}_{FV}}$-influence [Eq.~\eqref{eq:influkappa}] and (b) traditional $\kappa$-influence functions [Eq.~\eqref{eq:influkappaPRE}] plot for typical $\kappa$-values by considering. The black curve represents the classical influence function [Eq.~\eqref{eq:influclass}].}
    \label{fig:KappaGaussianInfluenceFunction}
\end{figure}

Panels (a) and (b) of Fig.~\ref{fig:KappaGaussianInfluenceFunction} show the $\kappa_{{}_{FV}}$- and traditional $\kappa$-influence functions plots, respectively, for typical $\kappa$-values. Note that large errors are neglected both in our proposal [Fig.~\ref{fig:KappaGaussianInfluenceFunction}(a)] and in the traditional $\kappa$-framework [Fig.~\ref{fig:KappaGaussianInfluenceFunction}(b)], and up-weighted in the classical case [black curve in Fig.~\ref{fig:KappaGaussianInfluenceFunction}], as already foreseen in the influence functions [Eqs.~\eqref{eq:influkappa}, \eqref{eq:influkappaPRE} and \eqref{eq:influclass}]. On the other hand, the small errors analysis reveals that both: the traditional $\kappa$-approach  and the classical framework [$\kappa \rightarrow 0$] show a similar behavior (linear weighting), as depicted in Fig.~\ref{fig:KappaGaussianInfluenceFunction}(b). In contrast, our proposal up-weight the small errors, especially in the limit case $\kappa \rightarrow 2/3$. In summary, the fact of constraining the second moment to the $\kappa$-Gaussian distribution was able to yield interesting features for the $\kappa_{{}_{FV}}$-objective function, such as the high-weighting small errors whilst providing a low-weighting of large errors, especially for values close to $2/3$ [see green curve in Fig.~\ref{fig:KappaGaussianInfluenceFunction}(a)]. 

%%%%%%%%%%%%%%%%%%%%%%%%%%%%%%%%%%%%%%%%%%%%%%%%%%%%%%%%%%%%%%%%%%%%%%%%%%%%%%
%%%%%%%%%%%%%%%%%%%%%%%%%%%%%%%%%%%%%%%%%%%%%%%%%%%%%%%%%%%%%%%%%%%%%%%%%%%%%%
%%%%%%%%%%%%%%%%%%%%%%%%%%%%%%%%%%%%%%%%%%%%%%%%%%%%%%%%%%%%%%%%%%%%%%%%%%%%%%
\section{\label{sec:NumericalExample} Numerical example} 
%%%%%%%%%%%%%%%%%%%%%%%%%%%%%%%%%%%%%%%%%%%%%%%%%%%%%%%%%%%%%%%%%%%%%%%%%%%%%%
%%%%%%%%%%%%%%%%%%%%%%%%%%%%%%%%%%%%%%%%%%%%%%%%%%%%%%%%%%%%%%%%%%%%%%%%%%%%%%
%%%%%%%%%%%%%%%%%%%%%%%%%%%%%%%%%%%%%%%%%%%%%%%%%%%%%%%%%%%%%%%%%%%%%%%%%%%%%%

In this section, with the intention of analyzing the robustness of the parameter estimates based on $\kappa$-Gaussian distribution, we consider a classic geophysical imaging problem known as Post-Stack Inversion (PSI) \cite{Russel1988}. The main objective of the PSI is to estimate the acoustic impedance of the subsurface (model parameters) from the seismic data (observed data) \cite{Ghosh_PSI_2000}. The constitutive relationship that associates the model space with the data space (that is, the forward problem) is given by the following direct operation \cite{Russel1988}:
\begin{equation}
    \textbf{d}_{mod} = \mathbf{S} \mathbf{D} \mathbf{m}, 
    \label{eq:modellededatacompletadef_matrix}
\end{equation}
where $\textbf{d}_{mod} \in \textbf{R}^{(c+j-1) \times u}$ is the modeled data, $\textbf{S} \in \textbf{R}^{(c+j-1) \times (j-1)}$ is a Toeplitz matrix computed from the seismic source $\textbf{s} = \{s_1, s_2, \cdots , s_c\} \in \textbf{R}^{c \times 1}$:
\begin{equation}
    \textbf{S} = %
   \begin{bmatrix}%
        s_1	&		0		&	\cdots    &	0\\
        s_2	&		s_1		&	\ddots    &	0\\
        \vdots		&	s_2	&	\ddots  &	0\\	
        s_c	&	\vdots		&   \ddots  &	s_1 \\
        0			& s_c		&  \vdots	&	s_2\\
        \vdots		&  \ddots		&  \ddots 	& \vdots \\
        0			& \cdots			&   0 	& s_c
    \end{bmatrix}
    ,
    \label{eq:wavelet_matrix}
\end{equation}
$\textbf{D} \in \textbf{R}^{(j-1) \times j}$ is defined as the following operator:
\begin{equation}
    \textbf{D} = \frac{1}{2}%
   \begin{bmatrix}%
        -1	&		\textcolor{white}{-}1 & 0		&\cdots &	0 	  \\
        \textcolor{white}{-}0		&	-1	&	1 & \ddots &\vdots\\	
        \textcolor{white}{-}\vdots		&  \ddots	& \ddots & \ddots	&  0 \\
        \textcolor{white}{-}0			&  \cdots			&   0 	& -1 & 1
    \end{bmatrix}
    ,
    \label{eq:wavelet_matrix}
\end{equation}
and $\textbf{m}$ is the logarithm of acoustic impedance. In addition, it is worth emphasizing that estimating the logarithm of the acoustic impedance is equivalent to estimating the acoustic impedance. In this way, the PSI forward operator is defined as: $\textbf{G} = \textbf{S} \textbf{D}$, and therefore the residual data is defined by $\varepsilon(\textbf{m}) = \textbf{S}\textbf{D}\textbf{m} - \textbf{d}^{obs}$,
where $\textbf{S}\textbf{D}\textbf{m}$ and $\textbf{d}^{obs}$ are the modeled and the observed data, respectively.

\begin{figure}[!htb]
\flushleft{(a) \hspace{5.4cm} (b) \hspace{5.5cm} (c)}
\resizebox{\textwidth}{!}{
    \includegraphics{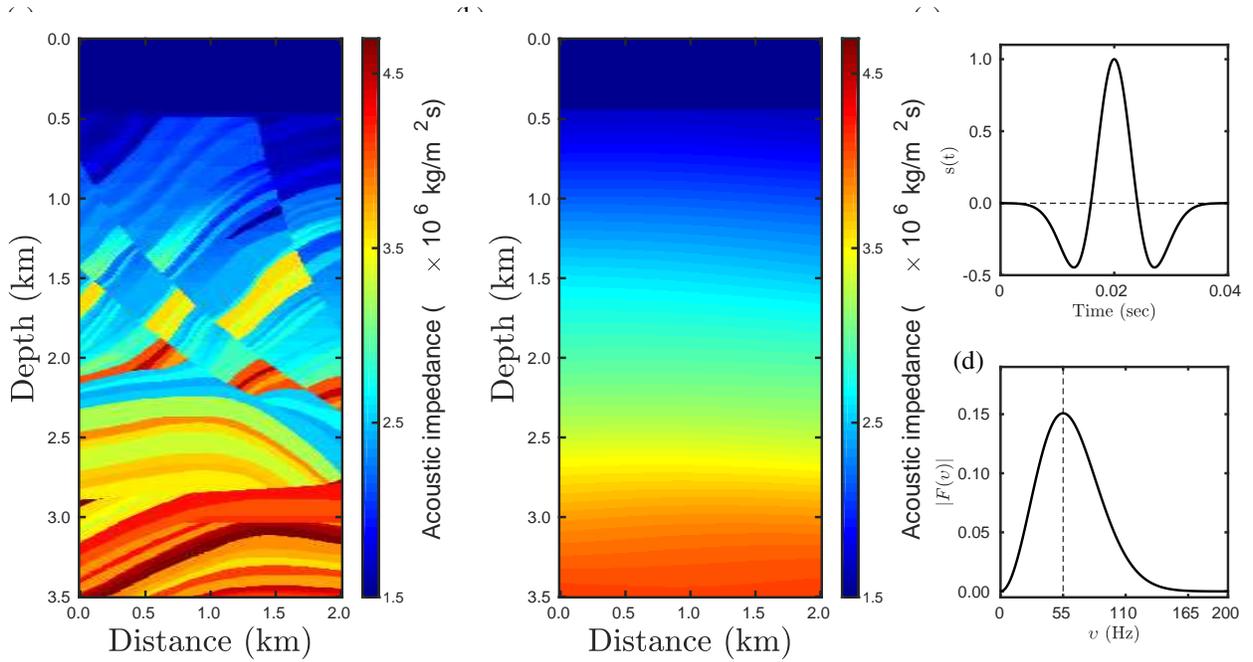}
    }
    \flushleft{\vspace{-4.4cm} \hspace{12.5cm} (d) \vspace{4cm}}
    \caption{(a) Marmousi2 realistic model used as the ground truth model (true model). (b)  Initial model for the PSI problem. (c) 55Hz Ricker wavelet used in this study as the seismic source [Eq.~\eqref{eq:ricker}], and its (d) amplitude spectrum $|F(\upsilon)|$.}
    \label{fig:true_and_initial_models}
\end{figure}

In all numerical simulations, we consider a section of a realistic geological model known as Marmousi2 model \cite{marmousi,Versteeg_marmousi} as the ground truth model (true model) [see Fig.~\ref{fig:true_and_initial_models}(a)]. It is worth emphasizing that this model is widely used to perform several tests of new seismic imaging strategies \cite{GeneticAlgoritm_Mazzotti_2016,FishWilton_PlosOne2019,IgoTsallisEntropy2020}. Furthermore, in all simulations carried out in this work, we consider a Ricker wavelet \cite{rickerSource} as the seismic source which is mathematically defined as:
\begin{equation}
        s(t)=\Big(1-2\pi^2\upsilon_p^2t^2\Big) \exp\Big(-\pi^2\upsilon_p^2t^2\Big),
        \label{eq:ricker}
\end{equation}
where $\upsilon_p$ represents the peak frequency (more energetic frequency content) and $t$ denotes the time. In the present study, we consider a peak frequency of $\upsilon_p = 55Hz$ as depicted in panels (c) and (d) of Fig.~\ref{fig:true_and_initial_models}.

We consider the \textit{limited}-\textit{memory} Broyden-Fletcher-Goldfarb-Shanno (\textit{l}-BFGS) method \cite{ByrdNocedalDetails,nocedal} to solve the optimization problems formulated in Eqs.~\eqref{eq:classicalobjectivefunction} and \eqref{eq:kappaObjectiveFunction}. The \textit{l}-BFGS algorithm updates the model parameters iteratively using the following relationship:
\begin{equation}
    \textbf{m}_{l+1}=\textbf{m}_{l}-\alpha_{l} \mathbf{H}^{-1}_l \nabla_m \phi(\textbf{m}_l),
    \label{eq:metodoquasiNewton}
\end{equation}
where $l = 0, 1, ..., N_{iter}$ with $N_{iter}$ being the maximum number of \textit{l}-BGFS iterations, and $\alpha > 0$ is the step-length which is computed according to the Wolfe conditions [see Ref.~\cite{wolfeOriginal} for more details]. $\nabla_m \phi(\textbf{m}_l)$ and $\mathbf{H}$ are the first- and second-order partial derivatives of the objective function, $\phi$, with respect to each model parameter. In this regard,  $\nabla_m \phi(\textbf{m}_l)$ and $\mathbf{H}^{-1}$ are the gradient of the objective function and the inverse of the Hessian matrix, respectively. In addition, it is worth highlighting that we have used the same initial model, $\textbf{m}_0$, [Fig.~\ref{fig:true_and_initial_models}(b)] for all simulations carried out in this work with $N_{iter} = 100$.

\subsection{Sensitivity to uncertainties in seismic source}

\begin{figure}[!htb]
\flushleft{(a) \hspace{8cm} (b)}
\resizebox{\textwidth}{!}{
    \includegraphics{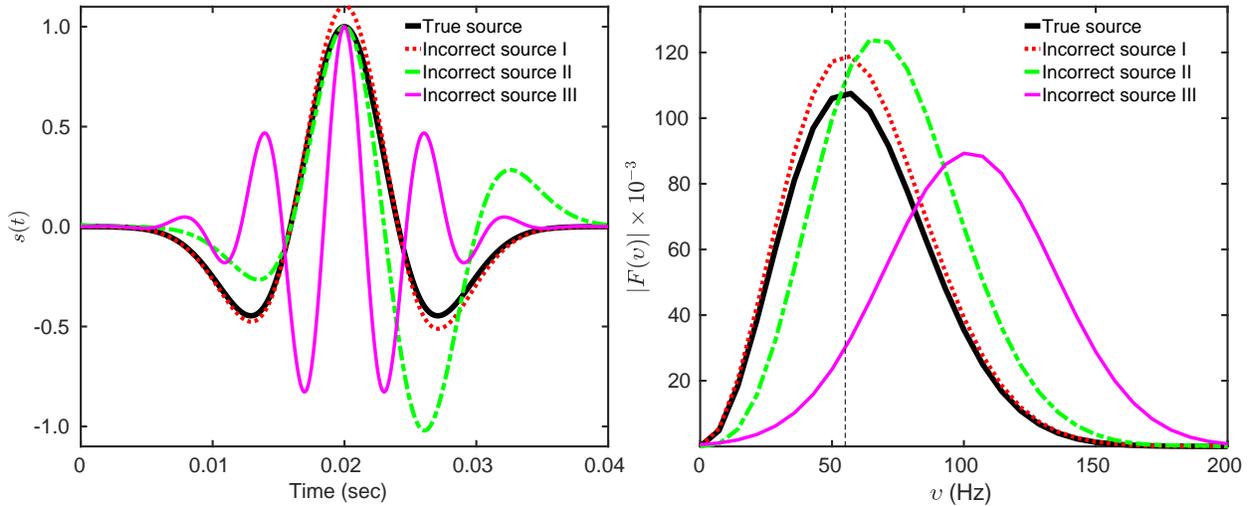}
    }
    \caption{Seismic sources used in the sensibility test regarding the incorrectly estimated sources, in which the exact source signature is depicted by the solid black line.}
    \label{fig:seismicSources}
\end{figure}

In geophysical applications, the exact source signature employed in a seismic survey is usually unknown, which is a hamper to perform reliable seismic inversions. In this way, we present in this section a numerical study to analyze the sensitivity of our proposal to inaccurate seismic sources. In this regard, we consider four different situations regarding the seismic source employed in the forward problem [Eq.~\eqref{eq:modellededatacompletadef_matrix}]. First, we consider the ideal case in which the seismic source corresponds to the exact source [see solid black line in Fig.~\ref{fig:seismicSources}]. In the last three ones, we consider incorrect sources named, respectively,  Incorrect source I, Incorrect source II, and Incorrect source III.

The Incorrect source I is a deformed version of the Ricker wavelet, which is given by: $s_1(t) = s(t) exp(5t)$, where $s(t)$ is defined in Eq.~\eqref{eq:ricker}. The Incorrect source II corresponds to the first temporal derivative of the Incorrect source I, $s_2(t) \propto \frac{d s_1(t)}{dt}$, normalized shortly thereafter. Finally, the Incorrect source III is known as Morlet wavelet of the form: $s_3(t) =exp(-t^2/2) cos(5t)$. The seismic sources signatures and their respective amplitude spectrum are depicted in Fig.~\ref{fig:seismicSources}. It is worth noting that in the present numerical experiments we consider noiseless-data, since our goal is to study exclusively the effect of the incorrectly estimated seismic source.

\begin{figure}[!htb]
\flushleft{(a) \hspace{3.2cm} (b) \hspace{3.2cm} (c) \hspace{3.3cm} (d)}
\resizebox{\textwidth}{!}{
    \includegraphics{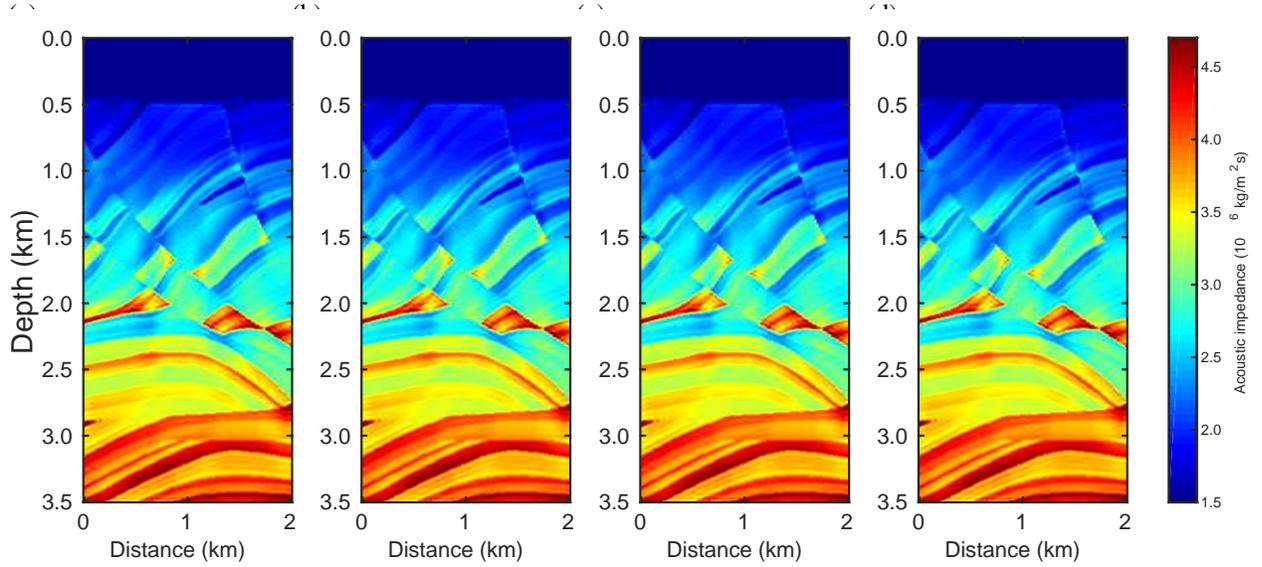}
    }
    \caption{Reconstructed acoustic impedance model, for the correct source case, using the PSI based on the (a) classical approach [Eq.~\eqref{eq:classicalobjectivefunction}], and the traditional $\kappa$-approach [Eq.~\eqref{eq:kappaObjectiveFunction} with $\beta_\kappa$ = 1/2], with (b) $\kappa = 0.1$, (c) $\kappa = 0.5$, and (d) $\kappa = 0.9$.}
    \label{fig:PSIResutls_PRE_CorrectSource}
\end{figure}

\begin{figure}[!htb]
\flushleft{(a) \hspace{3.2cm} (b) \hspace{3.2cm} (c) \hspace{3.3cm} (d)}
\resizebox{\textwidth}{!}{
    \includegraphics{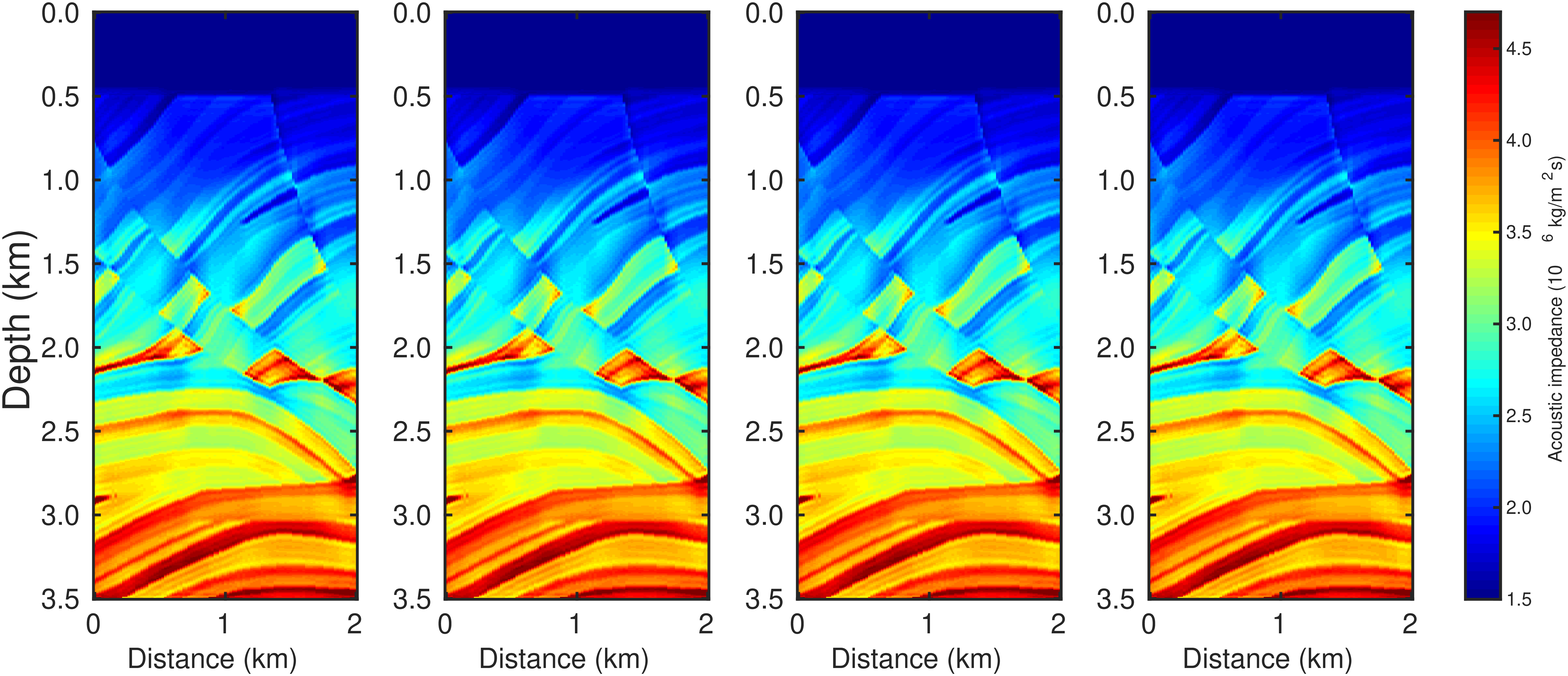}
    }
    \caption{Reconstructed acoustic impedance model, for the correct source case, using the PSI based on our proposal [Eq.~\eqref{eq:kappaObjectiveFunction}], with (a) $\kappa = 0.1$, (b) $\kappa = 0.3$, (c) $\kappa = 0.5$, and (d) $\kappa \rightarrow 2/3$.}
    \label{fig:PSIResutls_CorrectSource}
\end{figure}

\begin{figure}[!htb]
\flushleft{(a) \hspace{3.2cm} (b) \hspace{3.2cm} (c) \hspace{3.3cm} (d)}
\resizebox{\textwidth}{!}{
    \includegraphics{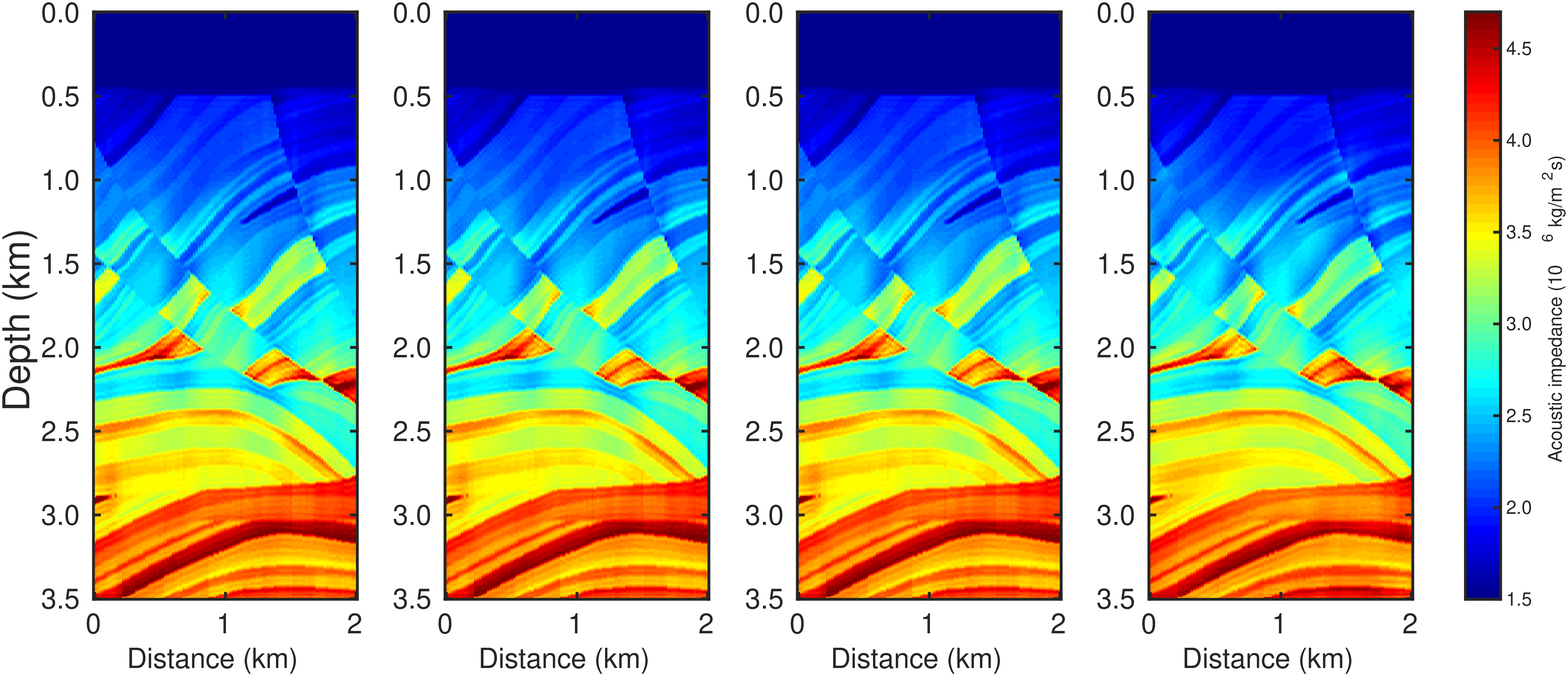}
    }
    \caption{Reconstructed acoustic impedance model, for the incorrect source I case, using the PSI based on the (a) classical approach [Eq.~\eqref{eq:classicalobjectivefunction}], and the traditional $\kappa$-approach [Eq.~\eqref{eq:kappaObjectiveFunction} with $\beta_\kappa$ = 1/2], with (b) $\kappa = 0.1$, (c) $\kappa = 0.5$, and (d) $\kappa = 0.9$.}
    \label{fig:PSIResutls_PRE_InCorrectSource1}
\end{figure}

\begin{figure}[!htb]
\flushleft{(a) \hspace{3.2cm} (b) \hspace{3.2cm} (c) \hspace{3.3cm} (d)}
\resizebox{\textwidth}{!}{
    \includegraphics{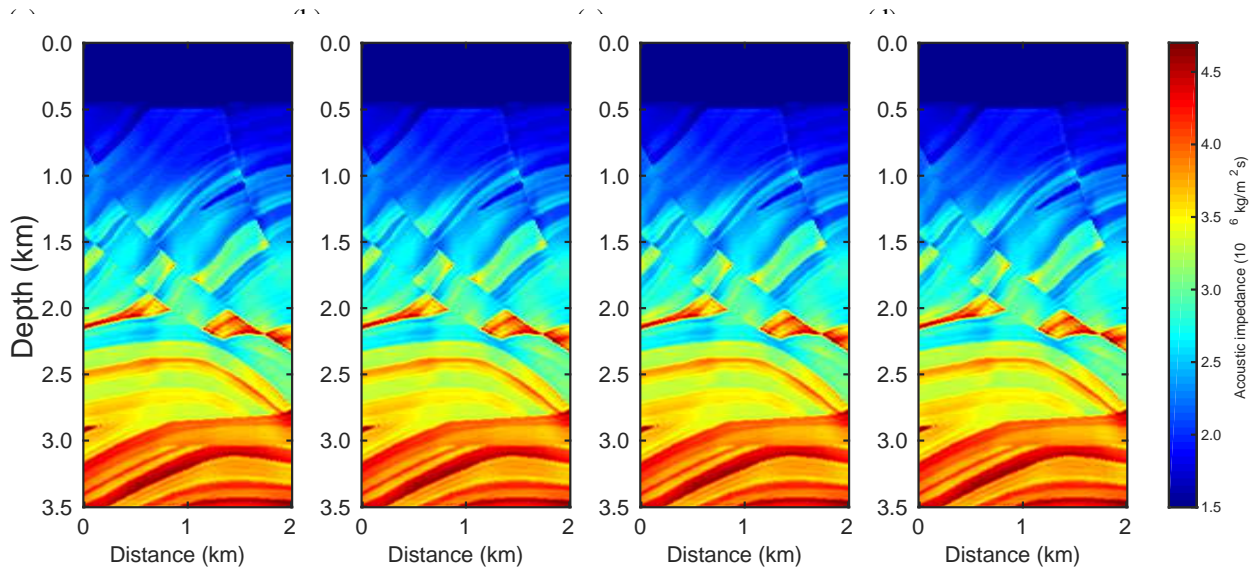}
    }
    \caption{Reconstructed acoustic impedance model, for the incorrect source I case, using the PSI based on our proposal [Eq.~\eqref{eq:kappaObjectiveFunction}], with (a) $\kappa = 0.1$, (b) $\kappa = 0.3$, (c) $\kappa = 0.5$, and (d) $\kappa \rightarrow 2/3$.}
    \label{fig:PSIResutls_InCorrectSource1}
\end{figure}

Figures~\ref{fig:PSIResutls_PRE_CorrectSource}-\ref{fig:PSIResutls_InCorrectSource1}  show the PSI results considering the ideal and Incorrect source I cases. From a visual inspection, we notice that the PSI results are satisfactory regardless of the objective function employed in the data-inversion, by reconstructing impedance models very close to the true model [Fig.~\ref{fig:true_and_initial_models}(a)]. Indeed, we include the ideal case in our numerical tests to validate the good functioning of the developed algorithms. In the case of the Incorrect source I, we conclude that small errors in the estimated seismic source are not a difficulty for the PSI problem.

\begin{figure}[!htb]
\flushleft{(a) \hspace{3.2cm} (b) \hspace{3.2cm} (c) \hspace{3.3cm} (d)}
\resizebox{\textwidth}{!}{
    \includegraphics{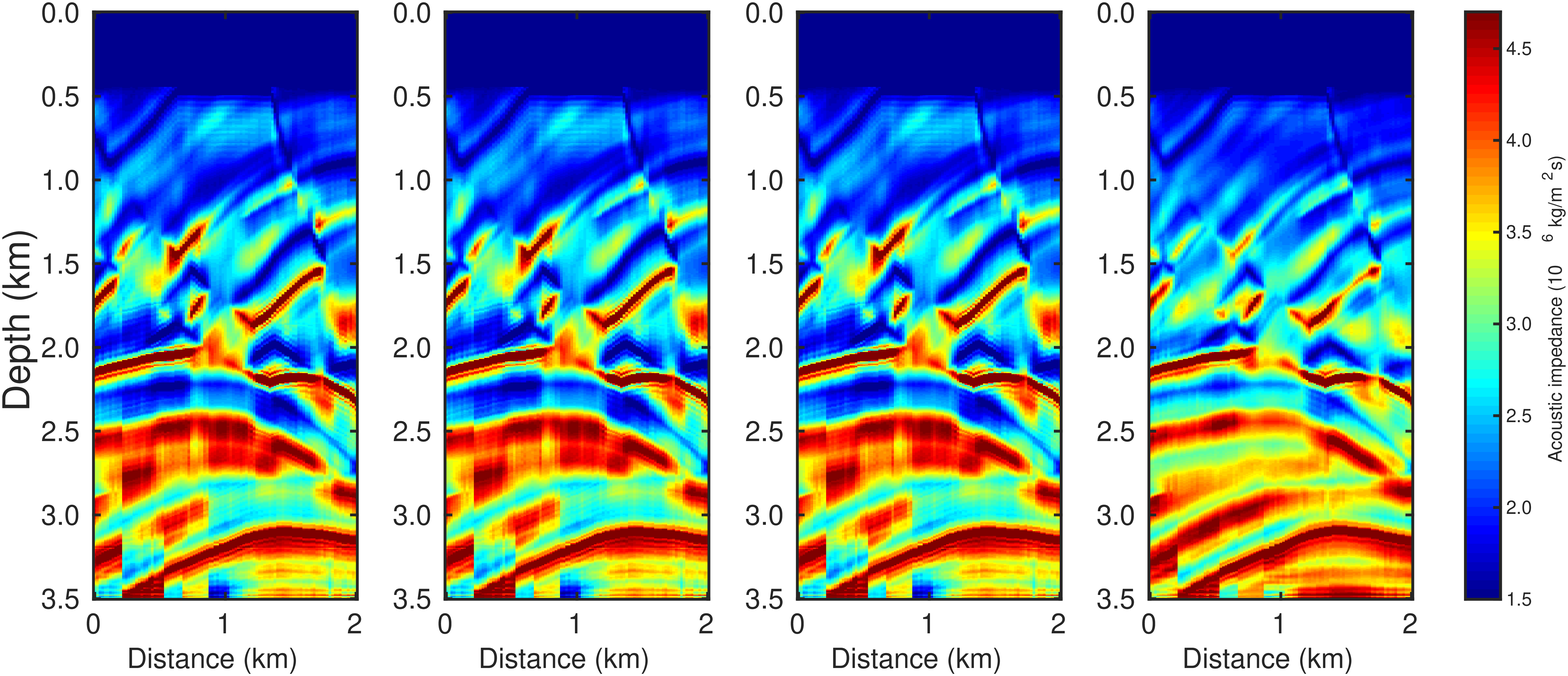}
    }
    \caption{Reconstructed acoustic impedance model, for the incorrect source II case, using the PSI based on the (a) classical approach [Eq.~\eqref{eq:classicalobjectivefunction}], and the traditional $\kappa$-approach [Eq.~\eqref{eq:kappaObjectiveFunction} with $\beta_\kappa$ = 1/2], with (b) $\kappa = 0.1$, (c) $\kappa = 0.5$, and (d) $\kappa = 0.9$.}
    \label{fig:PSIResutls_PRE_InCorrectSource2}
\end{figure}

\begin{figure}[!htb]
\flushleft{(a) \hspace{3.2cm} (b) \hspace{3.2cm} (c) \hspace{3.3cm} (d)}
\resizebox{\textwidth}{!}{
    \includegraphics{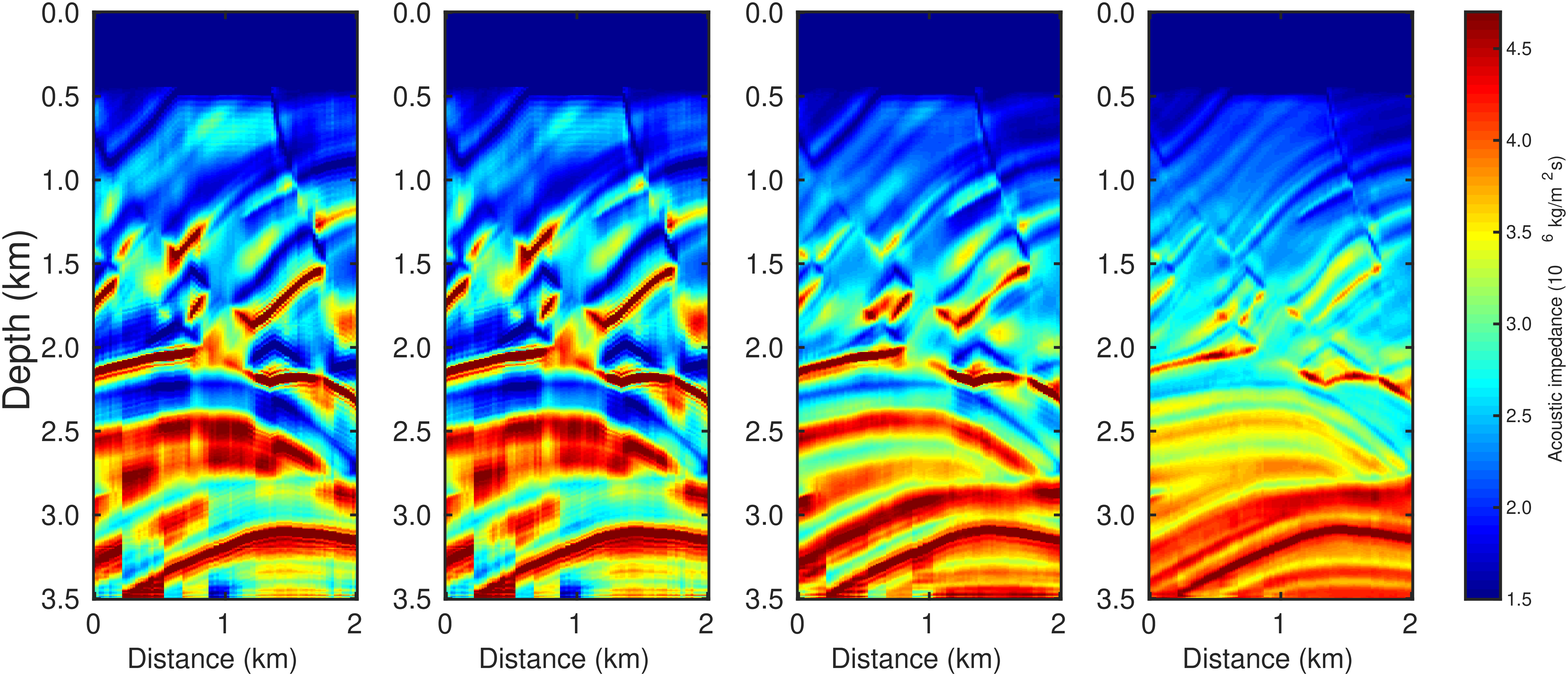}
    }
    \caption{Reconstructed acoustic impedance model, for the incorrect source II case, using the PSI based on our proposal [Eq.~\eqref{eq:kappaObjectiveFunction}], with (a) $\kappa = 0.1$, (b) $\kappa = 0.3$, (c) $\kappa = 0.5$, and (d) $\kappa \rightarrow 2/3$.}
    \label{fig:PSIResutls_InCorrectSource2}
\end{figure}

Regarding the reconstructed model obtained using the Incorrect source II, the PSI based on the conventional approach and on the traditional $\kappa$-approach  [Eq.~\eqref{eq:kappaObjectiveFunction}] failed to reconstruct acoustic impedance models, as depicted in Fig.~\ref{fig:PSIResutls_PRE_InCorrectSource2}, in which the impedance models have multiple reverberations and artifacts, as well as a low resolution when compared to the true model [Fig.~\ref{fig:true_and_initial_models}(a)]. In contrast, as the $\kappa$-value increases, our proposal reconstructs models closer to the true model, as depicted in Fig.~\ref{fig:PSIResutls_InCorrectSource2}. In fact, the accuracy of the reconstructed models based on our proposal is sensible to the $\kappa$-parameter, in which the $\kappa \rightarrow 2/3$ limit-case provides the more reliable estimated model [Fig.~\ref{fig:PSIResutls_InCorrectSource2}(d)].

\begin{figure}[!htb]
\flushleft{(a) \hspace{3.2cm} (b) \hspace{3.2cm} (c) \hspace{3.3cm} (d)}
\resizebox{\textwidth}{!}{
    \includegraphics{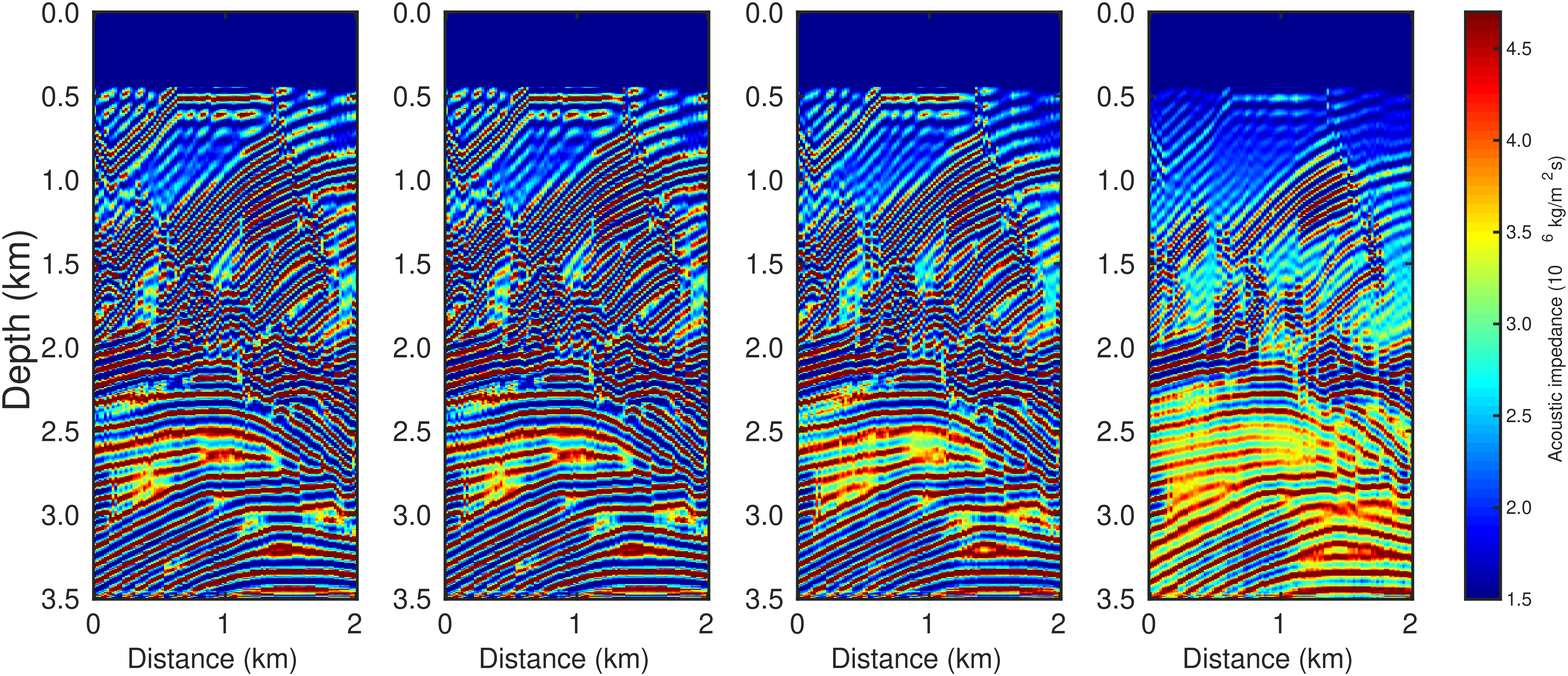}
    }
    \caption{Reconstructed acoustic impedance model, for the incorrect source III case, using the PSI based on the (a) classical approach [Eq.~\eqref{eq:classicalobjectivefunction}], and the traditional $\kappa$-approach [Eq.~\eqref{eq:kappaObjectiveFunction} with $\beta_\kappa$ = 1/2], with (b) $\kappa = 0.1$, (c) $\kappa = 0.5$, and (d) $\kappa = 0.9$.}
    \label{fig:PSIResutls_PRE_InCorrectSource3}
\end{figure}

\begin{figure}[!htb]
\flushleft{(a) \hspace{3.2cm} (b) \hspace{3.2cm} (c) \hspace{3.3cm} (d)}
\resizebox{\textwidth}{!}{
    \includegraphics{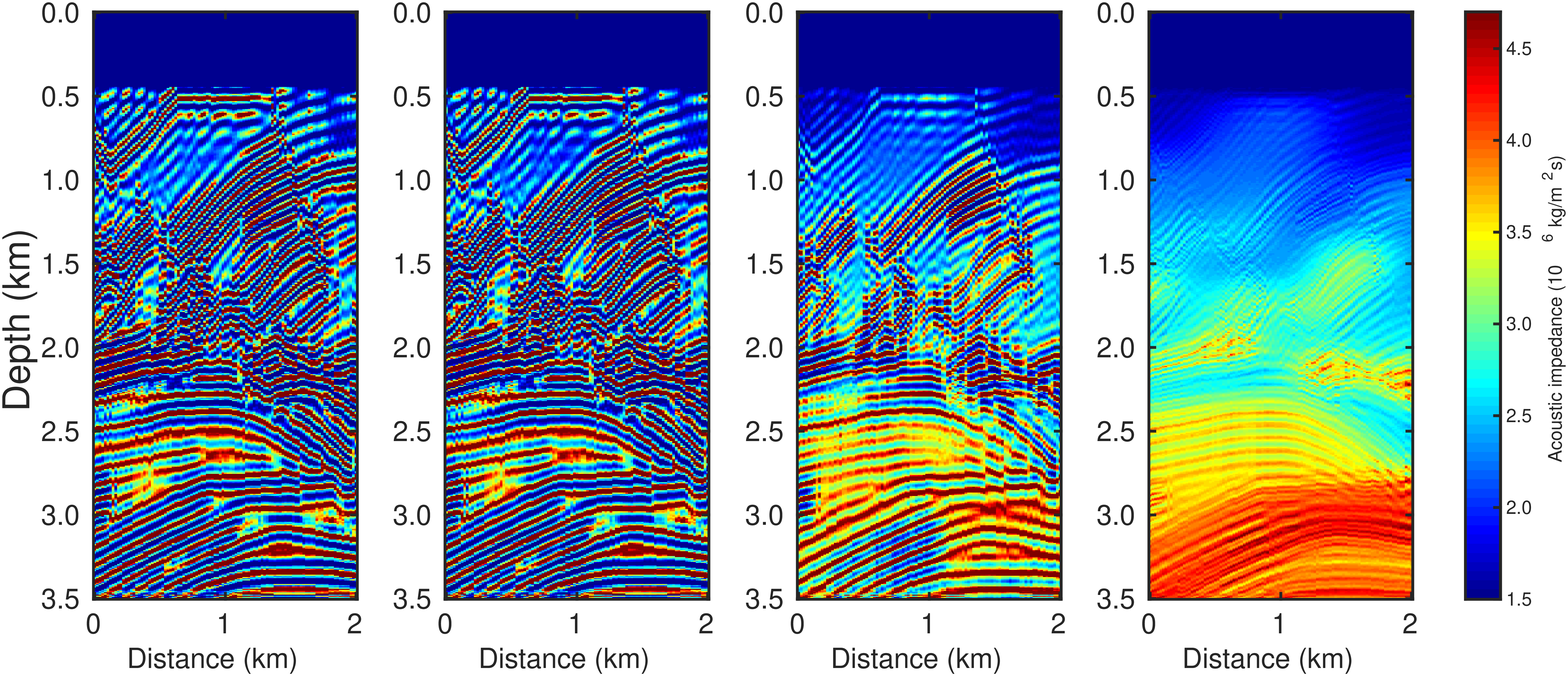}
    }
    \caption{Reconstructed acoustic impedance model, for the incorrect source III case, using the PSI based on our proposal [Eq.~\eqref{eq:kappaObjectiveFunction}], with (a) $\kappa = 0.1$, (b) $\kappa = 0.3$, (c) $\kappa = 0.5$, and (d) $\kappa \rightarrow 2/3$.}
    \label{fig:PSIResutls_InCorrectSource3}
\end{figure}

Now, in the Incorrect source III case, we notice that all approaches failed completely [Figs.~\ref{fig:PSIResutls_PRE_InCorrectSource3} and \ref{fig:PSIResutls_InCorrectSource3}], although our proposal, in the $\kappa \rightarrow 2/3$ limit-case [Fig.~\ref{fig:PSIResutls_InCorrectSource3}(d)], generated a model with fewer artifacts than the other strategies.

\subsection{Sensitivity to outliers}

To examine the outlier-resistance of the PSI based on the $\kappa_{{}_{FV}}$-Gaussian distribution, we consider seismic data corrupted by a weak background noise consisting of Gaussian errors with a signal-to-noise ratio (SNR) of $40$dB and a set of spikes (outliers). It is worth mentioning that the \textit{SNR} is computed through the ratio between the root-mean-square amplitude of the waveforms and the noise. Regarding spikes, we consider $34$ different scenarios, in which each scenario is related to a predetermined amount of spikes in the observed seismic data. For example, in the first scenario, we consider that $1\%$ of the samples of the observed data have spikes [$\%Spike = 1$]. In the second scenario, $4\%$ of the samples are contaminated by spikes [$\%Spike = 4$], and so on, every $3\%$ to a total of $100\%$ contamination [$\%Spike = 100$] in the last one. We randomly add spikes by using a uniform constant, in which the observed data with Gaussian noise at \textit{k}-th spike-position is computed as follows: $d^{obs}_i= [d^{obs}_i]_{Gaussian_{ Noise}} + \alpha \times \beta$, where $\alpha \in [-2;2]$ and $\beta$ follows a standard normal distribution. 

\begin{figure}[!htb]
\flushleft{(a) \hspace{3.2cm} (b) \hspace{3.2cm} (c) \hspace{3.3cm} (d)}
\resizebox{\textwidth}{!}{
    \includegraphics{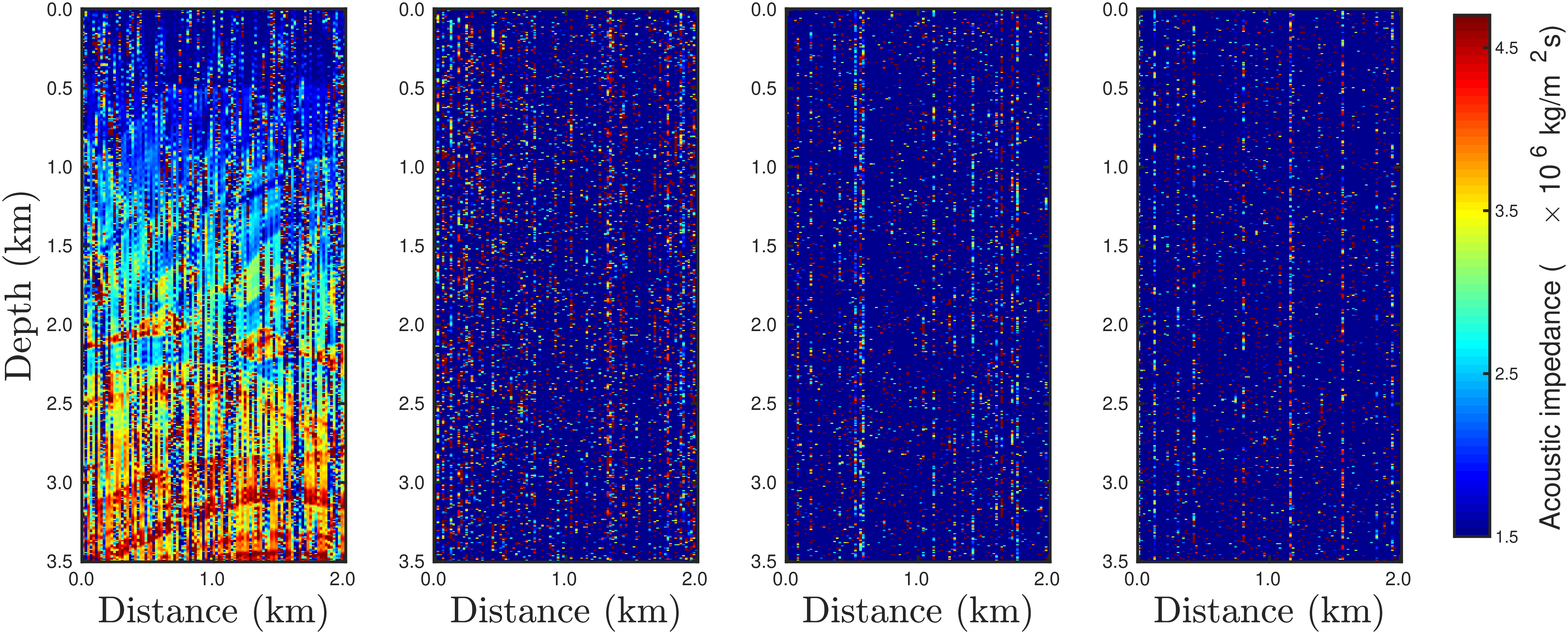}
    }
    \caption{Reconstructed acoustic impedance model using the PSI based on the classical approach [Eq.~\eqref{eq:classicalobjectivefunction}] for the cases: (a) $\%Spikes = 1\%$; (b) $\%Spikes = 31\%$; (c) $\%Spikes = 82\%$; and (d) $\%Spikes = 100\%$.}
    \label{fig:conventionalPSIResutls}
\end{figure}

\begin{figure}[!htb]
\flushleft{(a) \hspace{3.2cm} (b) \hspace{3.2cm} (c) \hspace{3.3cm} (d)}
\resizebox{\textwidth}{!}{
    \includegraphics{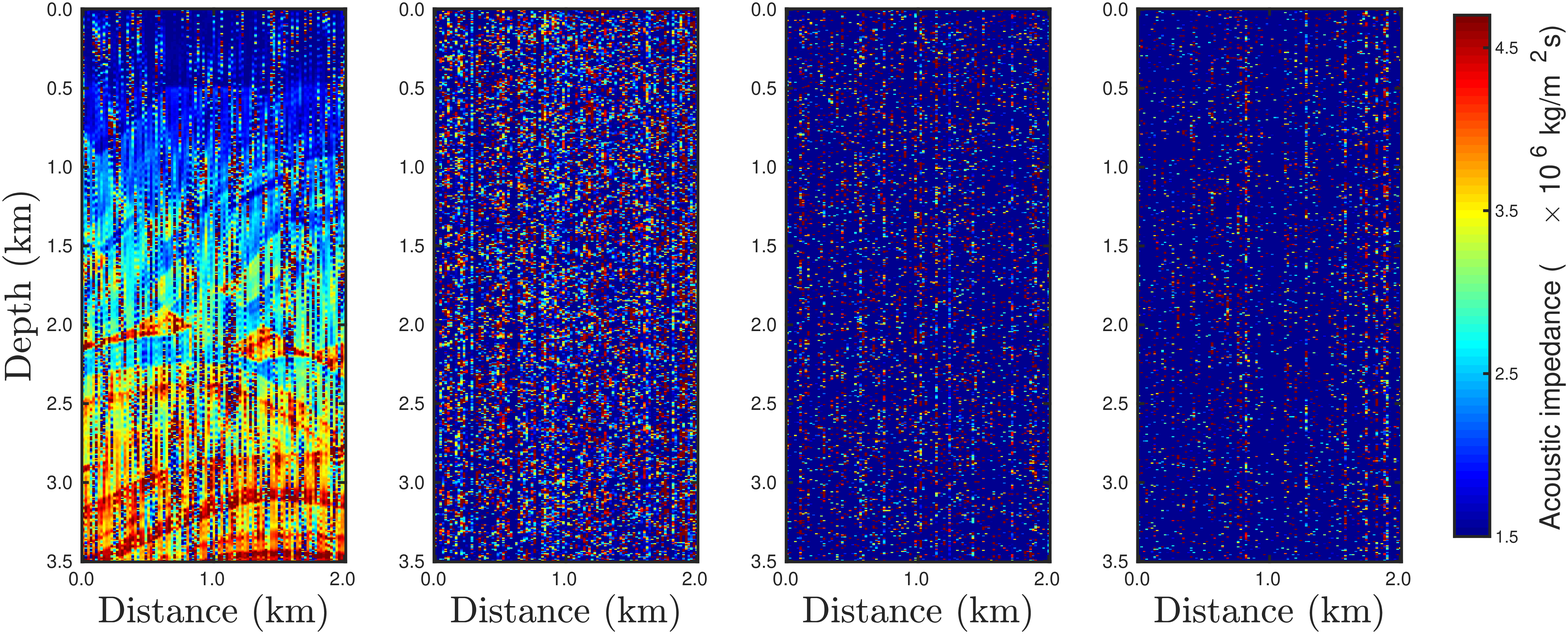}
    }
    \caption{Reconstructed acoustic impedance model using the PSI based on the traditional $\kappa$-approach [Eq.~\eqref{eq:kappaObjectiveFunction} with $\beta_\kappa$ = 1/2], with $\kappa = 0.9$, for the cases: (a) $\%Spikes = 1\%$; (b) $\%Spikes = 31\%$; (c) $\%Spikes = 82\%$; and (d) $\%Spikes = 100\%$.}
    \label{fig:PSIResutls_PRE_kappa_0_9}
\end{figure}

Figure~\ref{fig:conventionalPSIResutls} shows the reconstructed acoustic impedance model using the PSI based on the classical approach [Eq.~\eqref{eq:classicalobjectivefunction}] for spiky-noise scenarios $\%Spikes = 1\%$, $31\%$, $82\%$ and $100\%$. As already predicted by the classical influence function [Eq.~\eqref{eq:influclass}], only a $1\%$ contamination was enough for the classical approach to fail, as depicted in Fig.~\ref{fig:conventionalPSIResutls}(a). Indeed, the PSI based on the Gauss' law of error was strongly affected by the outliers in the seismic data set [panels (b)-(d) of Fig.~\ref{fig:conventionalPSIResutls}]. In a similar way, the PSI based on the traditional $\kappa$-approach  [Eq.~\eqref{eq:kappaObjectiveFunction} with $\beta_\kappa$ = 1/2] also failed to obtain an acoustic impedance good model, as also seen in Ref.~\cite{DASILVA2021125496}. In this regard, we carried out data inversions with $\kappa = 0.1$, $0.2$, $\cdots$, $0.9$, for each spike-noise-scenario.  Fig.~\ref{fig:PSIResutls_PRE_kappa_0_9} shows the best result regarding the approach proposed in Ref.~\cite{PhysRevE.101.053311_FWI_Kaniadakis} for the spiky-noise scenarios $\%Spikes = 1\%$, $31\%$, $82\%$ and $100\%$. 

\begin{figure}[!htb]
\flushleft{(a) \hspace{3.2cm} (b) \hspace{3.2cm} (c) \hspace{3.3cm} (d)}
\resizebox{\textwidth}{!}{
    \includegraphics{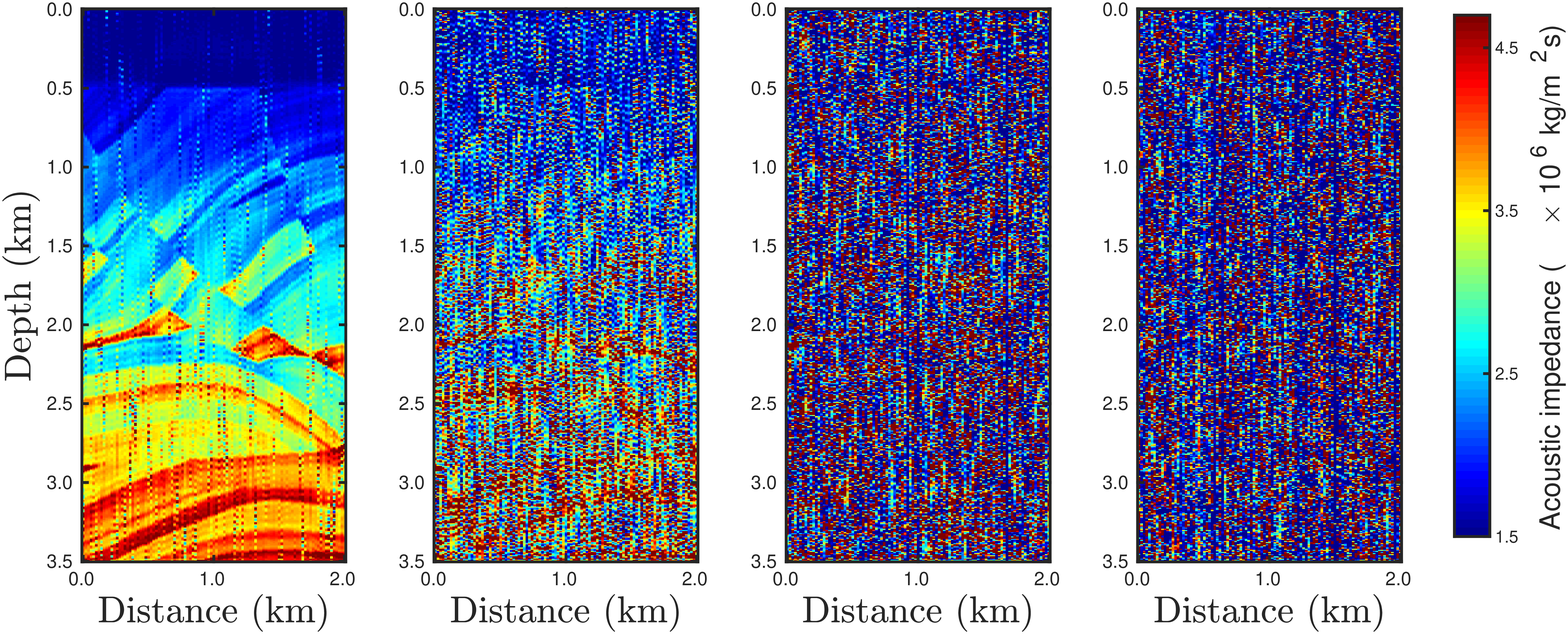}
    }
     \caption{Reconstructed acoustic impedance model using the PSI based on our proposal [Eq.~\eqref{eq:kappaObjectiveFunction}], with $\kappa = 0.66$, for the cases: (a) $\%Spikes = 1\%$; (b) $\%Spikes = 31\%$; (c) $\%Spikes = 82\%$; and (d) $\%Spikes = 100\%$.}
    \label{fig:PSIResutls_ourproposal_kappa_0_66}
\end{figure}

\begin{figure}[!htb]
\flushleft{(a) \hspace{3.2cm} (b) \hspace{3.2cm} (c) \hspace{3.3cm} (d)}
\resizebox{\textwidth}{!}{
    \includegraphics{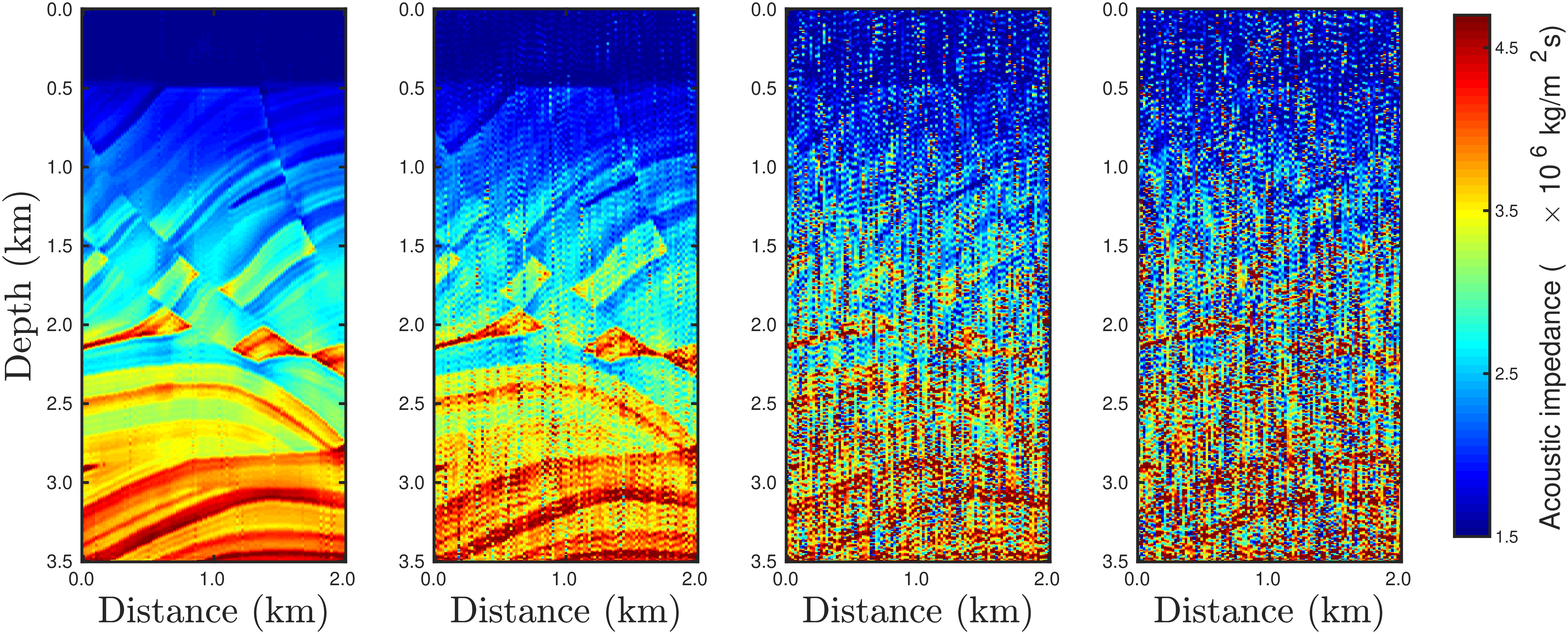}
    }
         \caption{Reconstructed acoustic impedance model using the PSI based on our proposal [Eq.~\eqref{eq:kappaObjectiveFunction}], with $\kappa = 0.666$, for the cases: (a) $\%Spikes = 1\%$; (b) $\%Spikes = 31\%$; (c) $\%Spikes = 82\%$; and (d) $\%Spikes = 100\%$.}
    \label{fig:PSIResutls_ourproposal_kappa_0_666}
\end{figure}

\begin{figure}[!htb]
\flushleft{(a) \hspace{3.2cm} (b) \hspace{3.2cm} (c) \hspace{3.3cm} (d)}
\resizebox{\textwidth}{!}{
    \includegraphics{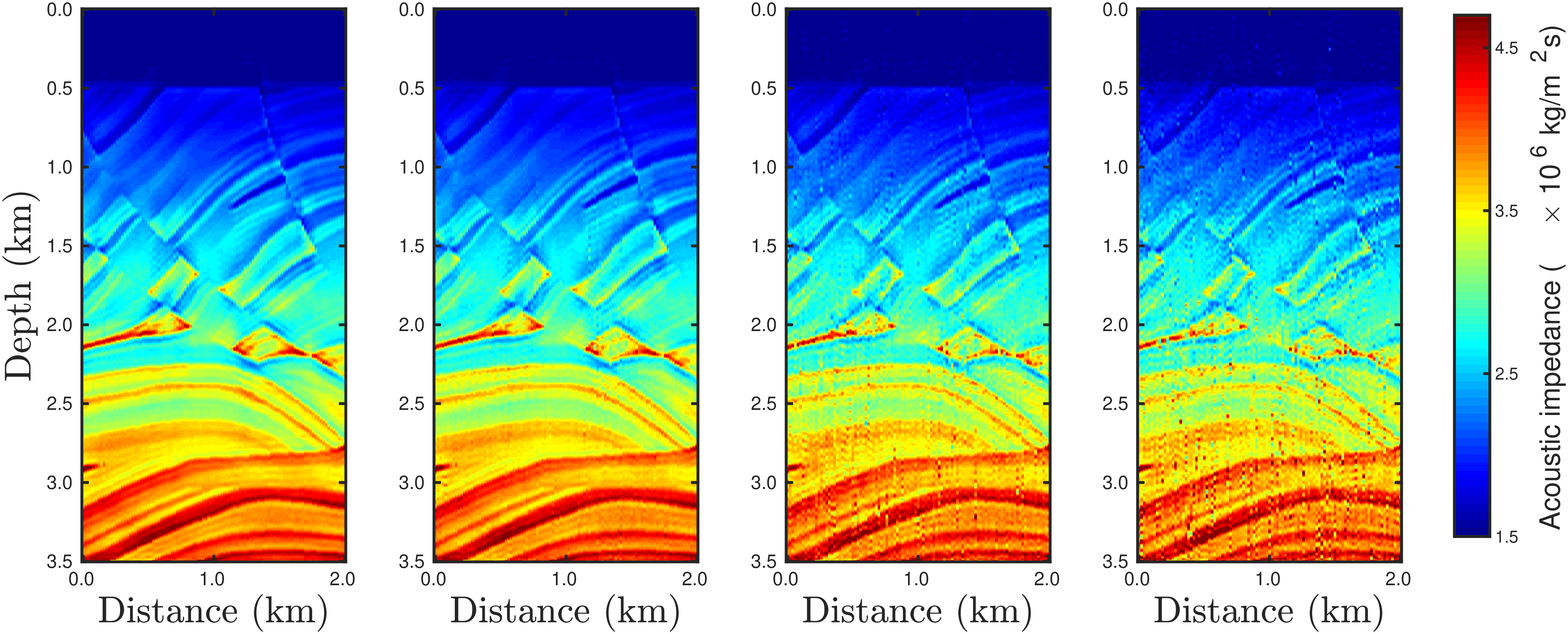}
    }
     \caption{Reconstructed acoustic impedance model using the PSI based on our proposal [Eq.~\eqref{eq:kappaObjectiveFunction}], with $\kappa = 0.6666$, for the cases: (a) $\%Spikes = 1\%$; (b) $\%Spikes = 31\%$; (c) $\%Spikes = 82\%$; and (d) $\%Spikes = 100\%$.}
    \label{fig:PSIResutls_ourproposal_kappa_0_6666}
\end{figure}

\begin{figure}[!htb]
\flushleft{(a) \hspace{3.2cm} (b) \hspace{3.2cm} (c) \hspace{3.3cm} (d)}
\resizebox{\textwidth}{!}{
    \includegraphics{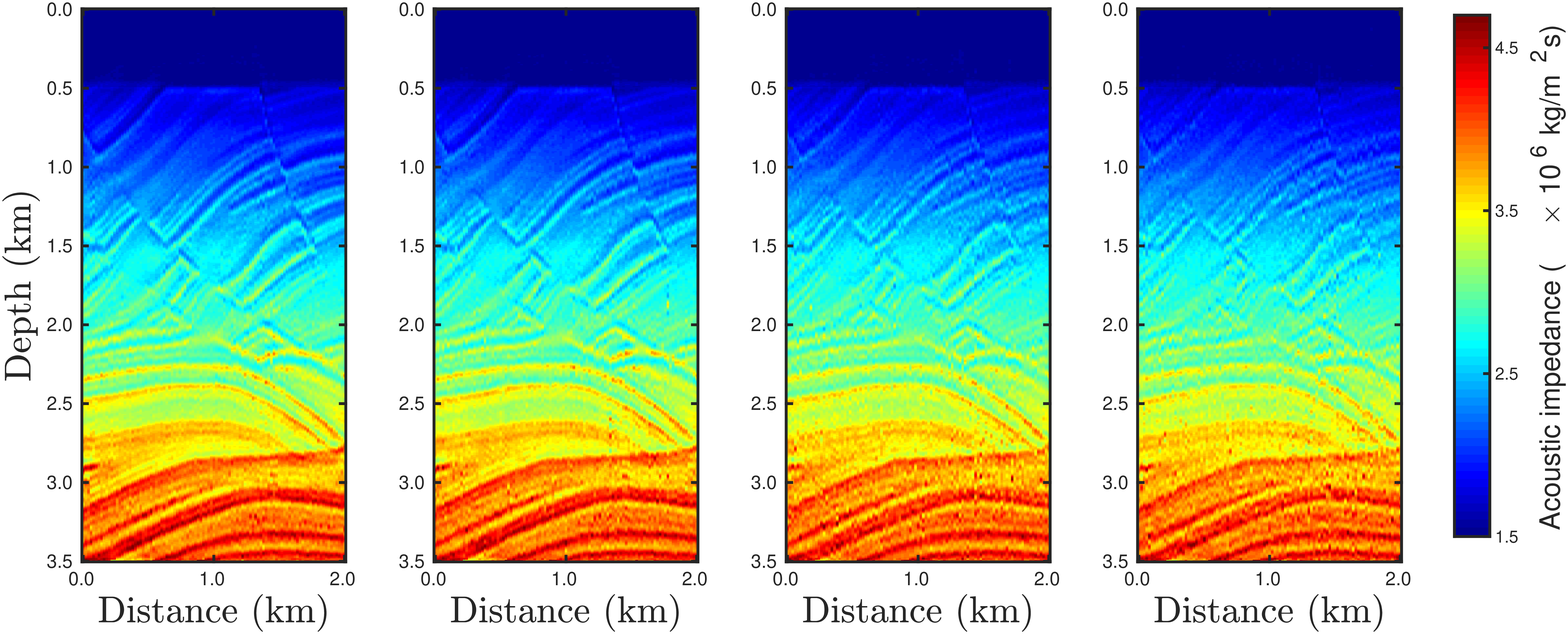}
    }
         \caption{Reconstructed acoustic impedance model using the PSI based on our proposal [Eq.~\eqref{eq:kappaObjectiveFunction}], with $\kappa = 0.666666666666667$, for the cases: (a) $\%Spikes = 1\%$; (b) $\%Spikes = 31\%$; (c) $\%Spikes = 82\%$; and (d) $\%Spikes = 100\%$.}
    \label{fig:PSIResutls_ourproposal_kappa_0_666666666666667}
\end{figure}

Regarding the PSI based on our proposal [Eq.~\eqref{eq:kappaObjectiveFunction}], we perform data inversion with $\kappa = 0.1$, $0.2$, $\cdots$, $0.6$, and $0.66$, $0.666$, $0.6666$, $\cdots$, $0.666666666666667$ for each spike-noise-circumstance. We notice that the inversion results for cases $\kappa \leq 0.66$ were all unsatisfactory and similar to that shown in Fig.~\ref{fig:PSIResutls_ourproposal_kappa_0_66},  with the exception of the case in which the percentage of spikes is low [see Fig.~\ref{fig:PSIResutls_ourproposal_kappa_0_66}(a)]. However, as the $\kappa$-value increases (even slightly), the reconstructed models have less artifacts (vertical lines) and, consequently, they are more close to the true model [Fig.~\ref{fig:true_and_initial_models}(a)], as depicted in Figs.~\ref{fig:PSIResutls_ourproposal_kappa_0_666}-\ref{fig:PSIResutls_ourproposal_kappa_0_666666666666667} for $\kappa = 0.6660$, $0.6666$ and $0.666666666666667$ cases, respectively. In these figures, note that we must choose the $\kappa$-value that is not so close to $2/3$ or so far. In fact, models rebuilt with $\kappa$-value away from $2/3$ are contaminated by many artifacts, especially when data quality is poor [see, for instance, Fig.~\ref{fig:PSIResutls_ourproposal_kappa_0_666}]. On the other hand, although inversions with a $\kappa$-value close to $2/3$ eliminate the artifacts observed in the cases shown above, they are not able to reconstruct the models with the high-resolution [see Fig.~\ref{fig:PSIResutls_ourproposal_kappa_0_666666666666667}] that case $\kappa = 0.6666$ [Fig.~\ref{fig:PSIResutls_ourproposal_kappa_0_6666}] achieved, for example.

\begin{figure}
\flushleft{(a) \hspace{7.5cm} (b)}
\resizebox{\textwidth}{!}{
    \includegraphics{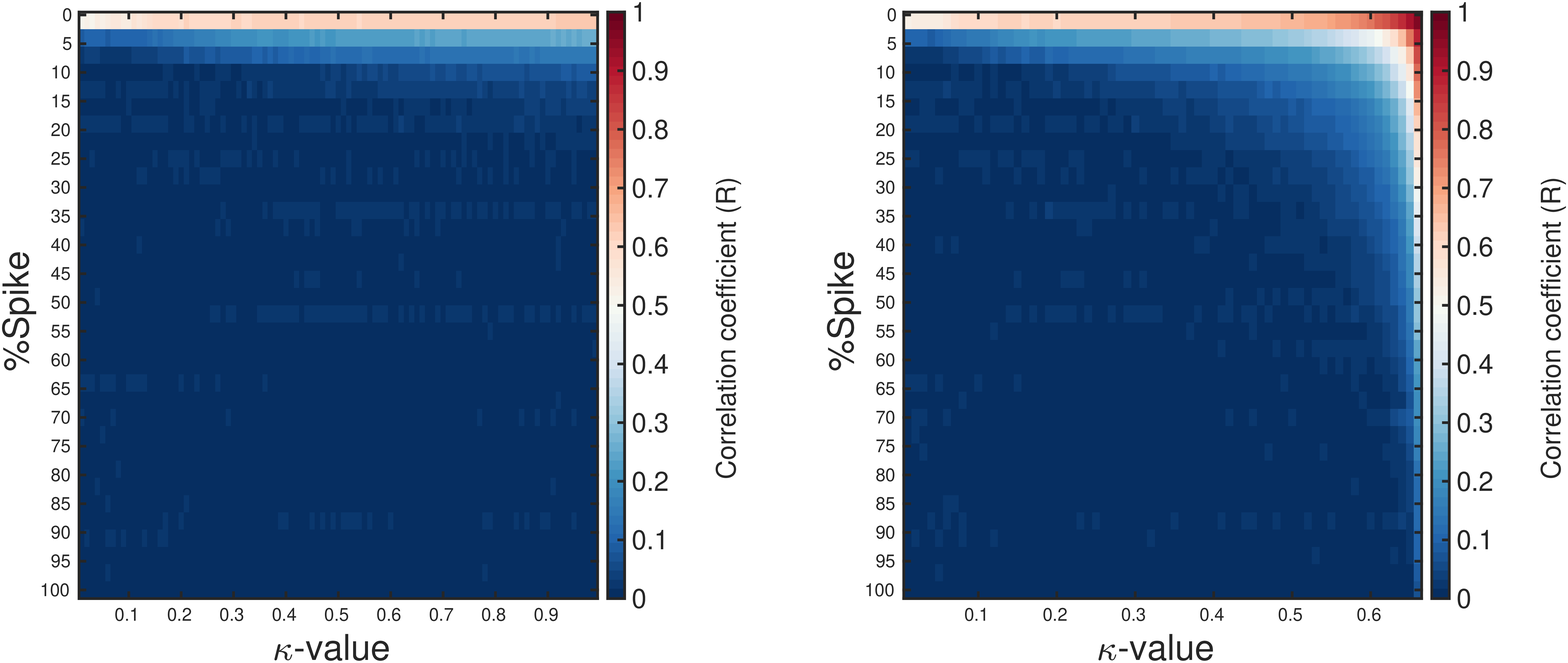}
    }
    \caption{Heatmaps created from Pearson's $R$ between the true model [Fig.~\ref{fig:true_and_initial_models}(a)] and the reconstructed models for all numerical experiment in the (a) PSI based on the traditional $\kappa$-approach  [Eq.~\eqref{eq:kappaObjectiveFunction} with $\beta_\kappa$ = 1/2], and in the (b) PSI based on our proposal [Eq.~\eqref{eq:kappaObjectiveFunction}].}
    \label{fig:heatMAP}
\end{figure}

In order to quantitatively compare all the models reconstructed in the present study, we consider the Pearson's correlation coefficient ($R$) \cite{pearson} to be the quality metric of the results. In this regard, we compute the Pearson’s $R$ between the true model [Fig.~\ref{fig:true_and_initial_models}(a)] and the reconstructed models. It is worth remembering that Pearson’s $R$ ranges between 0 (bad similarity)  and 1  (perfect similarity). We summarize the Pearson's $R$ values in a heatmap as depicted in Fig.~\ref{fig:heatMAP}, in which the panel (a) refers to the results of the PSI based on the traditional $\kappa$-approach [Eq.~\eqref{eq:kappaObjectiveFunction} with $\beta_\kappa$ = 1/2] and panel (b) refers to the results of the PSI based on our proposal [Eq.~\eqref{eq:kappaObjectiveFunction}]. The colors close to red (hot-colors) refer to the highest-correlations ($R \rightarrow 1$), whilst the colors close to blue (cold-colors) refer to the lowest-correlations. By a visual inspection of Fig.~\ref{fig:heatMAP}(a), we notice that both the conventional approach [$\kappa \rightarrow 0$] and the traditional $\kappa$-framework are only able to perform a reasonable data inversion in cases where the number of samples disturbed by spikes is close to $1\%$ [see the reddish region at the top of Fig.~\ref{fig:heatMAP}(a)]. In contrast, our proposal proves to perform an efficient data inversion and to be resistant to outliers for the $\kappa \rightarrow 2/3$ limit-case, as discussed in Section
~\ref{sec:metodologia}.

\begin{figure}[!htb]
\flushleft{(a) \hspace{3.7cm} (b) \hspace{3.5cm} (c) \hspace{3.5cm} (d)}
\resizebox{\textwidth}{!}{
    \includegraphics{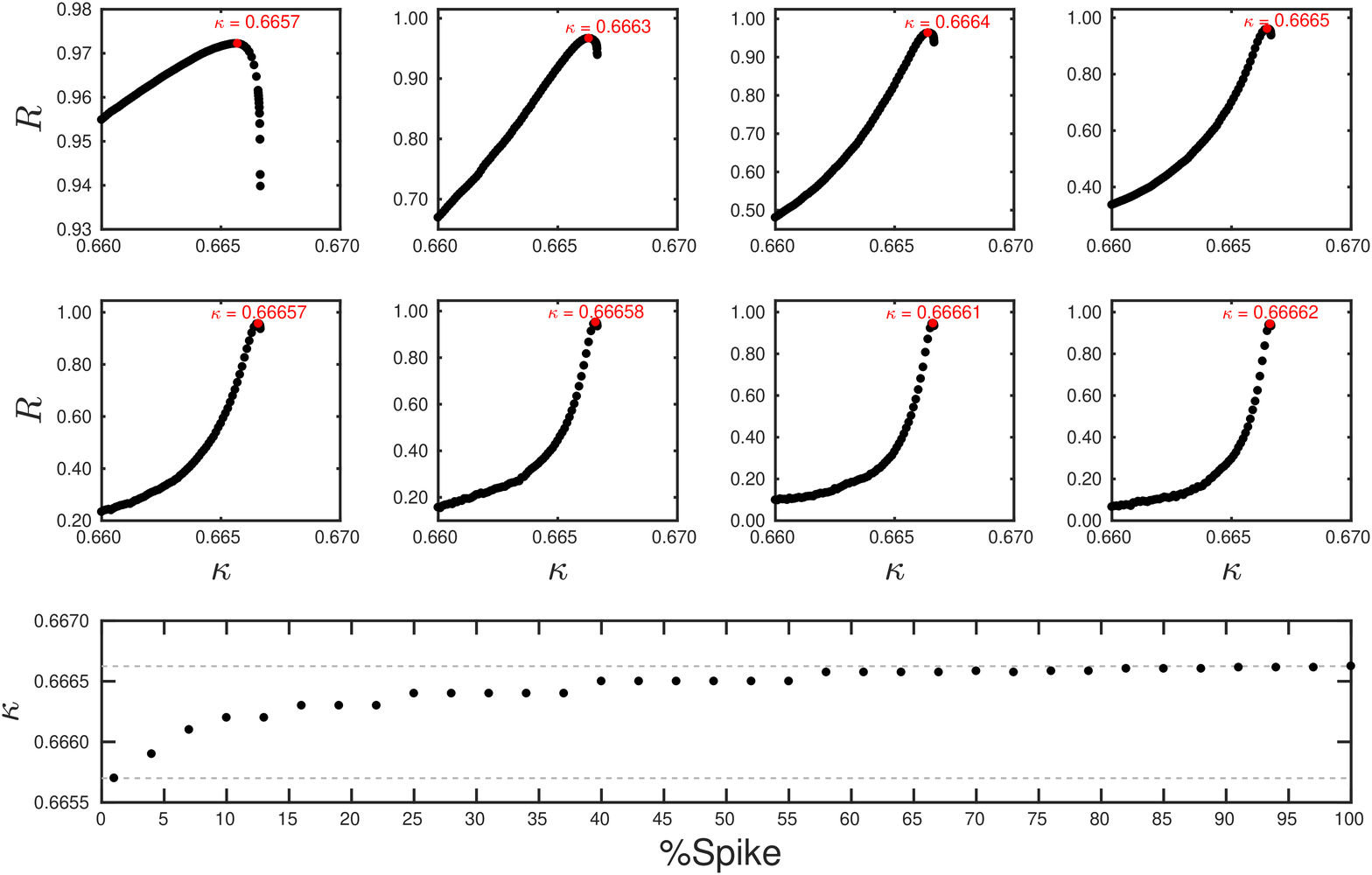}
    }
    \flushleft{\vspace{-7.5cm} (e) \hspace{3.7cm} (f) \hspace{3.5cm} (g) \hspace{3.6cm} (h)}
    \flushleft{\vspace{3cm} (i) \vspace{3.5cm}}
    \caption{Entropic $\kappa$-indexes associated with the PSI results based on our proposal [Eq.~\eqref{eq:kappaObjectiveFunction}] for the $0.66 < \kappa < 2/3$ case. The red dot indicates the $\kappa$-value that reconstructed the impedance acoustic model with the highest correlation ($R \rightarrow 1$) with the true model [Fig.~\ref{fig:true_and_initial_models}(a)] for the cases (a) $\%Spike = 1 \%$, (b) $\%Spike = 16 \%$, (c) $\%Spike = 31 \%$, (d) $\%Spike = 46 \%$, (e) $\%Spike = 61 \%$, (f) $\%Spike = 76 \%$, (g) $\%Spike = 91 \%$, and (h) $\%Spike = 100 \%$. (i) The $\kappa$-value associated with the reconstructed models with the highest correlations for all numerical experiments carried out in this study [$0.6657 \leq \kappa \leq 0.6667$].}
    \label{fig:kappaotimo}
\end{figure}

In fact, the $\kappa \rightarrow 2/3$ limit-case have a robustness property as discussed around Eq.~\eqref{eq:upsilon23} and by the numerical experiment presented at the end of Section~\ref{sec:metodologia}. In this way, we run more simulations for $\kappa$-values close to $2/3$, in order to check the existence of an optimal $\kappa$-value. Indeed, similarly to what was found in the previous session, there is an optimal $\kappa$-value, which are represented by the red dots in the panels (a)-(h) of Fig.~\ref{fig:kappaotimo} for the cases $\%Spike = 1 \%$, $16 \%$, $ 31 \%$, $46 \%$, $= 61 \%$, $76 \%$, $91 \%$, and $100 \%$, respectively. Analyzing all the numerical simulations, we verify that the optimal $\kappa$-value is in the $0.6657 \leq \kappa \leq 0.6667$ interval, in ascending order in relation to the data quality, as depicted in Fig.~\ref{fig:kappaotimo}(i).

%%%%%%%%%%%%%%%%%%%%%%%%%%%%%%%%%%%%%%%%%%%%%%%%%%%%%%%%%%%%%%%%%%%%%%%%%%%%%%
%%%%%%%%%%%%%%%%%%%%%%%%%%%%%%%%%%%%%%%%%%%%%%%%%%%%%%%%%%%%%%%%%%%%%%%%%%%%%%
%%%%%%%%%%%%%%%%%%%%%%%%%%%%%%%%%%%%%%%%%%%%%%%%%%%%%%%%%%%%%%%%%%%%%%%%%%%%%%
\section{\label{sec:finalremarks} Final Remarks} 
%%%%%%%%%%%%%%%%%%%%%%%%%%%%%%%%%%%%%%%%%%%%%%%%%%%%%%%%%%%%%%%%%%%%%%%%%%%%%%
%%%%%%%%%%%%%%%%%%%%%%%%%%%%%%%%%%%%%%%%%%%%%%%%%%%%%%%%%%%%%%%%%%%%%%%%%%%%%%
%%%%%%%%%%%%%%%%%%%%%%%%%%%%%%%%%%%%%%%%%%%%%%%%%%%%%%%%%%%%%%%%%%%%%%%%%%%%%%

In this work, by considering the necessity of an error law based on a probability distribution with finite variance, we have revisited the $\kappa$-generalized approach introduced in Ref.~\cite{PhysRevE.101.053311_FWI_Kaniadakis} to propose a new objective function (named $\kappa_{{}_{FV}}$-objective function) to mitigate the influence of spurious measures (outliers) in the context of inverse problem theory. From a statistical viewpoint, we explore the robustness properties of the $\kappa_{{}_{FV}}$-objective function by analyzing its influence function. In this regard, we have discussed the important role of the $\kappa$-parameter for robust inference of physical parameters from spiky noise in data from an analytical and numerical perspective.

Indeed, both the numerical experiment for estimating the mean of the normal distribution shown in Section~\ref{sec:metodologia} and the seismic data inversion results presented in Section~\ref{sec:NumericalExample}, proved that the inverse problem based on the $\kappa_{{}_{FV}}$-objective function is a powerful methodology for robust estimation of physical parameters even with very noisy data or uncertainties on the input parameter, which here was represented by the inaccurate seismic sources.. In particular, we have shown an optimal $\kappa$-value at limit $\kappa \rightarrow 2/3$, which makes the $\kappa_{{}_{FV}}$-objective function resistant to a lot of outliers. In this regard, by considering a classical geophysical problem, we have determined that the best data inversion results are related to $\kappa$-values in the range $0.6657 \leq \kappa \leq 0.6667$ [see Fig.~\ref{fig:kappaotimo}(i)].

The present results are significant because the data inversion based on $\kappa_{{}_{FV}}$-objective function ignores the outliers, which allows us to dispense a long pre-processing work before carrying out the data inversion. Our proposal is a promising tool also to be applied for describing a low-quality data set with many spikes. To conclude, it is worth emphasizing that the ideas introduced in the present work can be used to inference physical parameters, from measured data, in a wide range of several scientific and industrial fields.

\section*{Acknowledgments}
\label{sec:acknowledgements}
G.Z. dos Santos Lima, J.M. de Araújo and G. Corso gratefully acknowledge support from \textit{Petrobras} through the  project "\textit{Statistical physics inversion for multi-parameters in reservoir characterisation}" at Federal University of Rio Grande do Norte. R. Silva thanks \textit{Conselho Nacional de Desenvolvimento Científico e Tecnológico} (\textit{CNPq}) (Grant No. 307620/2019-0) for financial support. J.M. de Araújo thanks \textit{CNPq} for his productivity fellowship (Grant No. 313431/2018-3). G. Corso acknowledges \textit{CNPq} for support through productivity fellowship (Grant No. 307907/2019-8).

%Bibliography
%\bibliographystyle{unsrt}  
%\bibliography{references}  

\end{document}